\documentclass[11pt]{article}

\usepackage{microtype}
\usepackage{algorithm}
\usepackage{comment}
\usepackage[noend]{algorithmic}
\usepackage{xspace}
\usepackage{cleveref}
\usepackage{ifthen}
\usepackage{amsmath,amssymb,amsthm}
\usepackage{booktabs}
\usepackage{mleftright}
\usepackage{fullpage}

\usepackage{graphicx}
\usepackage{mathrsfs}
\usepackage[normalem]{ulem}
\usepackage{paralist}
\newcommand{\msg}[1]{\langle#1\rangle}
\renewcommand{\epsilon}{\varepsilon}

\newtheorem{definition}{Definition}

\newtheorem{theorem}{Theorem}
\newtheorem{lemma}{Lemma}

\newcommand{\I}{\mathcal{I}\xspace}
\newcommand{\G}{\mathcal{G}\xspace}
\newcommand{\cP}{\mathcal{P}\xspace}
\newcommand{\cA}{\mathcal{A}\xspace}

\newcommand{\findLOE}{\textsf{FindLightest}\xspace}
\newcommand{\controlledGHS}{\textsf{Controlled-GHS}\xspace}
\newcommand{\pipeline}{\textsf{Pipeline}\xspace}
\newcommand{\findPath}{\textsf{FindPath}\xspace}
\newcommand{\level}{\textsf{level}\xspace}
\newcommand{\computeMM}{{\sf ComputeMaximalMatching}\xspace}

\newcommand{\radius}{{\textsf{depth}}\xspace}

\let\originalleft\left
\let\originalright\right
\renewcommand{\left}{\mathopen{}\mathclose\bgroup\originalleft}
\renewcommand{\right}{\aftergroup\egroup\originalright}

\newcommand{\cover}{\textsf{ComputeCover}\xspace}
\newcommand{\up}{\textsf{up}\xspace}
\newcommand{\down}{\textsf{down}\xspace}
\newcommand{\cF}{{\mathcal{F}}\xspace}

\newcommand{\cM}{{\mathcal{M}}\xspace}
\newcommand{\cC}{{\mathcal{C}}\xspace}

\newcommand{\bGF}{\overline{G}_{\cF}\xspace}
\newcommand{\dGF}{\vec{G}_{\cF}\xspace}
\newcommand{\sk}[2]{\mathbf{s}_{#2}\langle#1\rangle\xspace}
\newcommand{\diam}{\textsf{diam}\xspace}
\newcommand{\bfs}{\textsf{bfs}\xspace}
\newcommand{\eff}{\mathcal{EF}\xspace}
\newcommand{\Geff}{{G}_{\eff}\xspace}

\newboolean{short}
\setboolean{short}{false}

\newcommand{\shortOnly}[1]{\ifthenelse{\boolean{short}}{#1}{}}
\newcommand{\onlyShort}[1]{\ifthenelse{\boolean{short}}{#1}{}}
\newcommand{\longOnly}[1]{\ifthenelse{\boolean{short}}{}{#1}}
\newcommand{\onlyLong}[1]{\ifthenelse{\boolean{short}}{}{#1}}

\def\polylog{\operatorname{polylog}}
\def\poly{\operatorname{poly}}

\begin{document}

\title{A Time- and Message-Optimal Distributed Algorithm for Minimum Spanning Trees\thanks{A preliminary
version of this paper~\cite{PanduranganRS17} appeared in the \emph{Proceedings of the 49th Annual ACM
Symposium on the Theory of Computing (STOC 2017).}}}
\author{
Gopal Pandurangan\thanks{Department of Computer Science, University of Houston, Houston, TX 77204, USA.
\hbox{E-mail}:~{\tt gopalpandurangan@gmail.com}.
Supported, in part, by NSF grants CCF-1527867, CCF-1540512, IIS-1633720, and CCF-1717075, and by US-Israel
Binational Science Foundation (BSF) grant 2016419.}
\and
Peter Robinson\thanks{Department of Computer Science, Royal Holloway, University of London, London, UK.
\hbox{E-mail}:~{\tt peter.robinson@rhul.ac.uk}.}
\and Michele Scquizzato\thanks{School of Computer Science and Communication, KTH Royal Institute of Technology, Stockholm, Sweden.
\hbox{E-mail}:~{\tt mscq@kth.se}. Supported, in part, by the European Research Council (ERC)
under the European Union's Horizon 2020 research and innovation programme under grant agreement No 715672.}
}

\maketitle

\begin{abstract}
This paper presents a randomized (Las Vegas) distributed algorithm that constructs a minimum spanning tree (MST) in
weighted networks with optimal (up to polylogarithmic factors) time and message complexity.
This algorithm  runs in $\tilde{O}(D + \sqrt{n})$ time and exchanges $\tilde{O}(m)$ messages (both with high probability), where
$n$ is the number of nodes of the network, $D$ is the diameter, and $m$ is the number of edges.
This is the \emph{first} distributed MST algorithm that matches
\emph{simultaneously} the time lower bound of $\tilde{\Omega}(D + \sqrt{n})$ [Elkin, SIAM J.\ Comput.\ 2006]
and the message lower bound of $\Omega(m)$ [Kutten et al., J.\ ACM 2015], which both apply to randomized Monte Carlo algorithms.

The prior time and message lower bounds are  derived using two
completely different graph constructions; the existing lower bound construction that shows one lower bound does not
work for the other.  To complement our algorithm, we  present
a new lower bound graph construction for which any distributed MST algorithm  requires both
$\tilde{\Omega}(D + \sqrt{n})$ rounds and $\Omega(m)$ messages.
\end{abstract}

%
%
%
%

\section{Introduction}\label{sec:intro}
The minimum-weight spanning tree (MST) construction problem is one of the central and most studied problems in distributed computing.
A long line of research aimed at developing efficient distributed algorithms for the MST problem started more than
thirty years ago with the seminal paper of Gallager, Humblet, and Spira~\cite{GallagerHS83}, which presented a
distributed algorithm that constructs an MST in $O(n \log n)$ rounds exchanging a total of $O(m + n \log n)$ messages\footnote{The
original algorithm has a message complexity of $O(m \log n)$, but it can be improved to $O(m + n \log n)$.}
(throughout, $n$ and $m$ will denote the number of nodes and the number of edges of the network, respectively).
The message complexity of this algorithm is (essentially) optimal,\footnote{It has been shown in~\cite{jacm15}
that the message complexity lower bound for leader election (and hence for any spanning tree construction as well) is $\Omega(m)$,
and this applies even to randomized Monte Carlo algorithms. On the other hand, it can be shown that an MST
can be constructed using $O(m)$ messages (but time
can be arbitrarily large) in any synchronous network~\cite{jacm15,PanduranganPS16}.} but its time complexity is not.
Hence further research concentrated on improving the time complexity. The time complexity
was first improved to $O(n \log \log n)$ by Chin and Ting~\cite{chin-almostlinear},
further improved to $O(n \log^* n)$ by Gafni~\cite{gafni-election}, and then to $O(n)$
by Awerbuch~\cite{awerbuch-optimal} (see also~\cite{faloutsos}).
The $O(n)$ bound is existentially optimal in the sense that there exist graphs for which this is the best possible.

This was the state of the art till the mid-nineties when Garay, Kutten, and Peleg~\cite{garay-sublinear}
raised the question of whether it is possible to identify graph parameters that can better 
capture the complexity of distributed network computations.
In fact, for many existing networks, their diameter\footnote{In this paper, by diameter we always mean unweighted diameter.}
$D$ is significantly smaller than the
number of vertices $n$, and therefore it is desirable to design protocols whose running
time is bounded in terms of $D$ rather than in terms of $n$.
Garay, Kutten, and Peleg~\cite{garay-sublinear} gave the
first such distributed algorithm for the MST problem with running
time $O(D + n^{0.614} \log^* n)$, which was later improved by Kutten and
Peleg~\cite{kutten-domset} to $O(D + \sqrt{n} \log^*n)$. However, both these algorithms
are not message-optimal,\footnote{In this paper, henceforth, when we say ``optimal'' we  mean ``optimal up to a $\polylog(n)$ factor''.}
as they exchange $O(m + n^{1.614})$ and $O(m + n^{1.5})$ messages, respectively. 
All the above results, as well as the one in this paper, hold in the  synchronous CONGEST model of distributed
computing, a well-studied standard model of distributed computing~\cite{peleg-locality} (see Section~\ref{sec:model}).

The lack of progress in improving the result of~\cite{kutten-domset}, and in particular breaking 
the $\tilde O(\sqrt{n})$ barrier,\footnote{$\tilde{O}(f(n))$ and
$\tilde{\Omega}(f(n))$ denote $O(f(n) \cdot \polylog(f(n)))$ and
$\Omega(f(n)/ \polylog(f(n)))$, respectively.} led to work on lower bounds for the distributed MST
problem. Peleg and Rubinovich~\cite{peleg-bound} showed that
$\Omega(D + \sqrt{n}/\log n)$ time is required by any distributed algorithm for constructing an MST,
even on networks of small diameter ($D = \Omega(\log n)$);  thus, this result
establishes the asymptotic near-tight optimality of the algorithm of~\cite{kutten-domset}.
The lower bound of Peleg and Rubinovich applies to exact, deterministic algorithms.
Later, the same lower bound of $\tilde{\Omega}(D + \sqrt{n})$ was shown for randomized (Monte Carlo)
and approximation algorithms as well~\cite{elkin,stoc11}. 

To summarize, the state of the art for distributed MST algorithms is that there exist algorithms which
are either time-optimal (i.e., they run in $\tilde{O}(D+ \sqrt{n})$ time) or message-optimal
(i.e., they exchange $\tilde{O}(m)$ messages), but not simultaneously both. Indeed, the time-optimal
algorithms of~\cite{kutten-domset,Elkin06}  (as well as the sublinear time algorithm of \cite{garay-sublinear}) are not
message-optimal, i.e., they require asymptotically much more than
$\Theta(m)$ messages. In contrast, the known message-optimal algorithms for MST
(in particular, \cite{GallagerHS83,awerbuch-optimal}) are not time-optimal, i.e., they take significantly more time than
$\tilde O(D+ \sqrt{n})$. 
In their 2000 SICOMP paper~\cite{peleg-bound}, Peleg and Rubinovich raised the question of 
whether one can design a distributed MST algorithm that is {\em simultaneously} optimal with respect to  time and message complexity.
In 2011, Kor, Korman, and Peleg~\cite{KorKP13} also raised this question and showed that
distributed {\em verification} of MST, i.e., verifying whether a given spanning tree is MST or not,
can be done in optimal messages and time, i.e., there exists a distributed verification algorithm
that uses $\tilde{O}(m)$ messages and runs in $\tilde{O}(D + \sqrt{n})$ time, and that these are optimal bounds for MST verification. 
However, the original question for MST construction remained open.

The above question addresses a fundamental aspect in distributed algorithms, namely the
relationship between the two basic complexity measures of time and messages.
The simultaneous optimization of both time and message complexity has been elusive
for several fundamental problems (including MST, shortest paths, and random walks),
and consequently research in the last three decades in distributed algorithms 
has focused mainly on optimizing either one of the two measures separately.  
However, in various modern and emerging applications such as 
resource-constrained communication networks  and distributed computation 
of large-scale data, it is crucial to design distributed algorithms 
that optimize  both measures {\em simultaneously}~\cite{soda15,podc15}. 

%
\subsection{Model and Definitions}\label{sec:model}
We first briefly describe the distributed computing model
in which our algorithm (as well as all the previously discussed MST
algorithms~\cite{chin-almostlinear,GallagerHS83,garay-sublinear,kutten-domset,awerbuch-optimal,gafni-election,Elkin06}) is specified and analyzed.
This is the CONGEST model (see, e.g., the book by Peleg~\cite{peleg-locality}), which is now standard in the distributed computing
literature.

A point-to-point communication network is modeled as an undirected weighted graph $G= (V,E, w)$,
where the vertices of $V$ represent the processors, the edges of $E$ represent the communication
links between them, and $w(e)$ is the weight of edge $e \in E$. Without loss of generality,
we assume that $G$ is connected. We also assume that the weights of the edges of
the graph are all distinct. This implies that the MST of the graph is unique.
The definitions and the results generalize readily to the case where the weights are not necessarily distinct.
Each node hosts a processor with limited initial knowledge. Specifically, we make the common
assumption that each node has unique identity numbers (this is not essential, but
simplifies presentation), and at the beginning of computation each
vertex $v$ accepts as input its own identity number and the weights
of the edges incident to it. Thus, a node has only {\em local}
knowledge.  Specifically we assume that each node has ports (each port having a unique port number); each incident edge is connected
to one distinct port. A node does not have any initial knowledge of the other endpoint of its incident edge (which node it is connected to or the port number that it is connected to). This model is  referred to as the {\em clean network model} in \cite{peleg-locality} and is also sometimes referred to as the
$KT_0$ model, i.e., the initial (K)nowledge of
all nodes is restricted (T)ill radius 0 (i.e., just the local knowledge) \cite{peleg-locality}. The $KT_0$ model is a standard model in distributed computing and typically used in the  literature (see e.g., \cite{peleg-locality,Tel,lynch,Attiya}), including  all the prior results on distributed MST  (e.g., ~\cite{awerbuch-optimal,chin-almostlinear,GallagerHS83,garay-sublinear,kutten-domset,gafni-election,Elkin06}) with a notable exception (\cite{KingKT15},
discussed in detail in Section~\ref{sec:related}).%

The vertices are
allowed to communicate through the edges of the graph $G$. It is assumed
that communication is synchronous and occurs in discrete rounds
(time steps). 
In each time step, each node $v$ can send an arbitrary message of
$O(\log n)$ bits through each edge $e = (v,u)$ incident to $v$,
and each message arrives at $u$ by the end of this time step.
(If unbounded-size messages are allowed---this is the so-called LOCAL
model---the MST problem can be trivially solved in $O(D)$ time~\cite{peleg-locality}.) 
The weights of the edges are at most
polynomial in the number of vertices $n$, and therefore the weight
of a single edge can be communicated in one time step. This model of
distributed computation is called the CONGEST$(\log n)$
model or simply the CONGEST model \cite{peleg-locality}. 

The efficiency of distributed algorithms is traditionally measured
by their time and message (or, communication) complexities. 
Time complexity measures the number of 
synchronous rounds taken by the algorithm, whereas
message complexity measures the total amount of messages 
sent and received by all the processors during the execution of the 
algorithm. Both complexity measures crucially influence the performance 
of a distributed algorithm.
We say that a problem enjoys {\em singular optimality} if it admits
a distributed algorithm whose time and message complexity are
both optimal. When the problem fails to admit such a solution, namely, 
algorithms with better time complexity for it necessarily incur higher 
message complexity and vice versa, we say that the problem exhibits a {\em time-message tradeoff}.

\subsection{Our Results}\label{sec:result}
\paragraph{Distributed MST Algorithm}  In this paper we present a distributed MST algorithm in the CONGEST model which is simultaneously 
time- and message-optimal. The algorithm is randomized Las Vegas, and always returns the MST. 
The running time of the algorithm is $\tilde{O}(D+ \sqrt{n})$ and the message complexity is
$\tilde{O}(m)$, and both bounds hold with high probability.\footnote{Throughout,
with high probability (w.h.p.) means with probability $\ge 1 -1/n^{\Omega(1)}$, where $n$ is
the network size.} This is the first distributed MST algorithm that matches
\emph{simultaneously} the time lower bound of $\tilde{\Omega}(D + \sqrt{n})$~\cite{elkin,stoc11}
and the message lower bound of $\Omega(m)$ \cite{jacm15}, which both apply even to randomized
Monte Carlo algorithms, thus closing a more than thirty-year-old line of research in distributed computing.
In terms of the terminology introduced earlier, we can therefore say that the distributed MST problem
exhibits singular optimality up to polylogarithmic factors. Table~\ref{table} summarizes the known
upper bounds on the complexity of distributed MST. 

\begin{table}[h!]
  \centering
  \begin{tabular}{lcc}
    \toprule
    Reference & Time Complexity & Message Complexity\\
    \midrule
    Gallager et al.~\cite{GallagerHS83} & $O(n \log n)$ & $O(m + n \log n)$\\
    Awerbuch~\cite{awerbuch-optimal} & $O(n)$ & $O(m + n \log n)$\\
    Garay et al.~\cite{garay-sublinear} & $O(D + n^{0.614}\log^*n)$ & $O(m + n^{1.614})$\\
    Kutten and Peleg~\cite{kutten-domset} & $O(D + \sqrt{n}\log^*n)$ & $O(m + n^{1.5})$\\
    Elkin~\cite{Elkin06} & $\tilde{O}(\mu(G,w) + \sqrt{n})$ & $O(m + n^{1.5})$\\
    This paper & $\tilde O(D + \sqrt{n})$ & $\tilde O(m)$\\    
    \bottomrule
  \end{tabular}\caption{Summary of upper bounds on the complexity of distributed MST. }\label{table}
\end{table}

\paragraph{Lower Bound} Both the aforementioned time and message lower bounds are existential, and are derived using two
completely different graph constructions. However, the graph used to show one lower bound {\em does not}
work for the other. %
To complement our main result, in Section~\ref{sec:lowerbound} we present
a new graph construction for which any distributed MST algorithm  requires \emph{both} 
$\tilde{\Omega}(D + \sqrt{n})$ rounds and $\Omega(m)$ messages.

\subsection{Other Related Work}\label{sec:related}
Given the importance of the distributed MST problem, there has been significant work over the last 30 years on this problem and related aspects.
Besides the prior work already mentioned in Section~\ref{sec:intro}, we now discuss other relevant work on distributed MST.

\paragraph{Other Distributed MST Algorithms}
Elkin~\cite{Elkin06} showed that a parameter called \emph{MST-radius} captures the
complexity of distributed MST algorithms better. The MST-radius, denoted by $\mu(G,w)$,
and which is a function of the graph topology as well as the edge weights,
roughly speaking is the maximum radius each vertex has to examine to check whether
any of its edges is in the MST. Elkin devised a distributed protocol
that constructs the MST in $\tilde{O}(\mu(G,w) + \sqrt{n})$ time.
The ratio between diameter and MST-radius can be
as large as $\Theta(n)$, and consequently, on some inputs, this
protocol is faster than the protocol of~\cite{kutten-domset} by a
factor of $\Omega(\sqrt{n})$. However, a drawback of this protocol
(unlike the previous MST protocols~\cite{kutten-domset,garay-sublinear,chin-almostlinear,gafni-election,GallagerHS83})
is that it cannot detect the termination of the algorithm in that
time (unless $\mu(G,w)$ is given as part of the input). 
On the other hand, it can be shown that for distributed MST algorithms that correctly terminate $\Omega(D)$ is a lower bound
on the running time \cite{peleg-bound,disc14}. (In fact, \cite{disc14} shows that
for every  sufficiently large $n$ and every function $D(n)$ with $2 \le D(n) < n/4$, there exists a graph $G$ of $n' \in \Theta(n)$ nodes and diameter $D' \in \Theta(D(n))$ which requires  $\Omega(D')$ rounds to compute a spanning tree with constant probability.)
We also note that the message complexity of Elkin's algorithm is $O(m + n^{3/2})$.

Some classes of graphs admit efficient MST algorithms that beat the general $\tilde{\Omega}(D + \sqrt{n})$
time lower bound. This is the case for planar graphs, graphs of bounded genus, treewidth, or
pathwidth~\cite{GhaffariH16,HaeuplerIZ16,HaeuplerIZ16a}, and graphs with small random walk
mixing time~\cite{GhaffariKS17}.

\paragraph{Time Complexity}
From a practical perspective,  given that MST construction can
take as much as $\Omega(\sqrt{n}/\log n)$ time even in low-diameter
networks, it is worth investigating whether one can
design distributed algorithms that run faster and output an
approximate minimum spanning tree.
The question of devising faster approximation algorithms for MST
was raised in \cite{peleg-bound}. Elkin~\cite{elkin} later
established a hardness result on distributed MST approximation,
showing that {\em approximating} the MST problem on a certain
family of graphs of small diameter (e.g., $O(\log n)$) within
a ratio $H$ requires essentially $\Omega(\sqrt{n/H \log n})$ time.
Khan and Pandurangan~\cite{disc} showed that there can be an exponential time gap between
exact and approximate MST construction by showing that there exist graphs where any distributed (exact) MST algorithm takes $\Omega(\sqrt{n}/\log n)$ rounds, whereas an $O(\log n)$-approximate MST can be computed in $O(\log n)$ rounds.
The distributed approximation algorithm of Khan and Pandurangan is
message-optimal but not time-optimal.

Das Sarma et al.~\cite{stoc11} 
settled the time complexity of distributed approximate MST by showing that 
this problem, as well as approximating shortest paths and about twenty other 
problems, satisfies a time lower bound of $\tilde{\Omega}(D + \sqrt{n})$. 
This applies to deterministic as well as randomized algorithms, and to
both exact and approximate versions. 
In other words, any distributed algorithm for computing a $H$-approximation 
to MST, for any $H > 0$, takes $\tilde{\Omega}(D + \sqrt{n})$ time in the worst case.

\paragraph{Message Complexity}
Kutten et al.~\cite{jacm15} fully settled the message complexity of leader election in general
graphs, even for randomized algorithms and under very general settings. Specifically, they showed that any randomized
algorithm (including Monte Carlo algorithms with suitably large constant success probability) requires  $\Omega(m)$  messages;
this lower bound  holds for any $n$ and $m$, i.e., given any $n$ and $m$, there exists a graph with $\Theta(n)$ nodes
and $\Theta(m)$ edges for which the lower bound applies.
Since a distributed MST algorithm can also be used to elect a leader (where the root of the tree is the leader,
which can be chosen using $O(n)$ messages once a tree is constructed), the
above lower bound applies to distributed MST construction as well, for all $m \geq cn$, where $c$ is a sufficiently large constant.
 
The above bound holds even for {\em non-comparison} algorithms, that is algorithms that may also manipulate the actual value of node's identities, not just  compare identities with each other, and even  if nodes have initial knowledge of $n, m$, and $D$. 
It also holds for synchronous networks, and even if all the nodes wake up simultaneously.
Finally, it holds not only for the CONGEST model~\cite{peleg-locality}, where sending a message of $O(\log n)$ bits takes one unit of time, but also for the LOCAL model~\cite{peleg-locality}, where the number of bits carried in a single message can be arbitrary.

\paragraph{The $KT_1$ Variant}
It is important to point out that this paper and all the prior results discussed above (including the prior MST results 
~\cite{awerbuch-optimal,chin-almostlinear,GallagerHS83,garay-sublinear,kutten-domset,gafni-election,Elkin06})
assume the so-called {\em clean network model}, a.k.a.\ $KT_0$~\cite{peleg-locality} (cf.\ Section \ref{sec:model}), where nodes
do not have initial knowledge of the identity of their neighbors.
However, one can assume a model where nodes do have such a knowledge.
This model is called the {\em $KT_1$ model}.
Although the distinction between $KT_0$ and $KT_1$ has clearly no bearing on the asymptotic bounds for the time complexity,
it is significant when considering message complexity.
Awerbuch et al. \cite{vainish} show that $\Omega(m)$ is a message lower bound for MST in the $KT_1$ model,  
if one allows only (possibly randomized Monte Carlo) comparison-based algorithms, i.e., algorithms that can operate on IDs only by comparing them.
(We note that all prior MST algorithms mentioned earlier are comparison-based, including ours.)  
Hence, the result of \cite{vainish} implies that our MST algorithm (which is comparison-based and randomized) is {\em time- and message-optimal}
in the $KT_1$ model if one considers comparison-based algorithms only.

Awerbuch et al.~\cite{vainish} also show that the $\Omega(m)$ message lower bound  applies even to non-comparison based (in particular,  algorithms that can perform arbitrary local computations)  {\em deterministic} algorithms 
in the CONGEST model  that terminate in a time bound that depends only on the graph topology (e.g., a function of $n$). 
On the other hand, for {\em randomized non-comparison-based} algorithms, it turns out that the message lower bound
of $\Omega(m)$ does not apply in the $KT_1$ model. 
Recently, King et al.~\cite{KingKT15} showed a surprising and elegant result: in the $KT_1$ model one can give a randomized Monte Carlo algorithm to construct an MST in $\tilde{O}(n)$ messages ($\Omega(n)$ is a message lower bound) and in $\tilde{O}(n)$ time. This algorithm is randomized and not comparison-based. While this algorithm shows that one can achieve $o(m)$ message complexity (when $m = \omega(n \polylog n)$), it is {\em not} time-optimal (it can take significantly more than $\tilde \Theta(D+\sqrt{n})$ rounds). In subsequent work, Mashreghi and King~\cite{MashreghiK17} presented
another randomized, not comparison-based MST algorithm with round complexity $\tilde{O}(\text{Diam(MST)})$ and with message complexity $\tilde{O}(n)$.
It is an open question whether one can design a randomized  (non-comparison based) algorithm that takes $\tilde{O}(D + \sqrt{n})$ time and $\tilde{O}(n)$ messages in the $KT_1$ model. 

\paragraph{Subsequent Work}  The preliminary version of this paper~\cite{PanduranganRS17} raised the open problem of
whether there exists a {\em deterministic} time- and message-optimal MST algorithm. We notice that our algorithm is {\em randomized},
due to the use of the randomized cover construction of~\cite{Elkin06}, even though the rest of the algorithm is deterministic.
Elkin~\cite{Elkin17}, building on our work, answered this question affirmatively by devising a deterministic MST algorithm that
achieves essentially the same bounds as in this paper, i.e., uses $\tilde{O}(m)$ messages and runs in $\tilde{O}(D+ \sqrt{n})$
time.\footnote{Actually, the bounds are better than in this paper by logarithmic factors.} Elkin's algorithm is simpler as it bypasses
Phase~2 of Part~2 of our algorithm, and thus bypasses the randomized cover construction; the rest of the high-level structure
of Elkin's algorithm is similar to our algorithm.

\section{High-Level Overview of the Algorithm}\label{sec:randalgo}

The time- and message-optimal distributed MST algorithm of this paper builds on prior distributed MST algorithms
that were either message-optimal or time-optimal but {\em not both}. We provide a high-level overview of our algorithm and  some intuition behind it;
we also compare and contrast it with  previous MST algorithms.
The full description of the algorithm and its analysis are given in Section~\ref{sec:detailalgo}.
The algorithm can be divided into two parts as explained next.

\subsection{First Part: $\controlledGHS$}\label{sec:part1}
We first run  the so-called $\controlledGHS$ algorithm,
which was first used in the sublinear-time distributed MST
algorithm of Garay, Kutten, and Peleg~\cite{garay-sublinear}, as well as in the time-optimal algorithm of Kutten and Peleg~\cite{kutten-domset}.
$\controlledGHS$ is the (synchronous version of the) classical Gallager-Humblet-Spira (GHS) algorithm~\cite{GallagerHS83,peleg-locality}, with some modifications.
We recall that the synchronous GHS algorithm, which is essentially a distributed implementation of Bor\r{u}vka's algorithm---see, e.g., \cite{peleg-locality},
consists of $O(\log n)$ phases. In the initial phase each node is an {\em MST fragment}, by which we mean a connected subgraph of the MST.
In each subsequent phase, every MST fragment finds a lightest (i.e., minimum-weight) outgoing edge (LOE)---these edges are guaranteed
to be in the MST by the  cut property~\cite{tarjan}. The MST fragments are merged via the LOEs to form larger MST fragments.
The number of phases is $O(\log n)$, since the number of MST fragments gets at least halved in each phase.
The message complexity is $O(m + n \log n)$, which is essentially optimal,
and the time complexity is $O(n \log n)$. The time complexity is not optimal because much of the communication during a
phase uses {\em only the MST fragment edges}. Since the diameter of an MST fragment can be as large as $\Omega(n)$
(and this can be significantly larger than the graph diameter $D$), the time complexity of the GHS algorithm is not optimal. 

The $\controlledGHS$ algorithm alleviates this situation by controlling the growth of the diameter of the MST fragments during merging.
At the end of $\controlledGHS$, at most $\sqrt{n}$ fragments remain, each of which has diameter $O(\sqrt{n})$.
These are called {\em base fragments}.
$\controlledGHS$ can be implemented using $\tilde{O}(m)$ messages in
$\tilde{O}(\sqrt{n})$ rounds. (Note that $\controlledGHS$ as implemented in the time-optimal algorithm
of~\cite{kutten-domset} is not message-optimal---the messages exchanged can be $\tilde{O}(m+n^{3/2})$;
however, a modified version can be implemented using  $\tilde{O}(m)$ messages, as explained in Section~\ref{sec:cghs}.)

\subsection{Second Part: Merging the $\sqrt{n}$ Remaining Fragments}
The second part of our algorithm, after the $\controlledGHS$ part, is different from the existing time-optimal MST algorithms.
The existing time-optimal MST algorithms~\cite{kutten-domset,Elkin06}, as well as the algorithm of~\cite{garay-sublinear},
are not message-optimal since they use the $\pipeline$ procedure of~\cite{peleg-time-optimal,garay-sublinear}.
The $\pipeline$ procedure builds an auxiliary breadth-first search (BFS) tree of the network, collects all the {\em inter-fragment} edges
(i.e., the edges between the $\sqrt{n}$ MST fragments) at the root
of the BFS tree, and then finds the MST locally. The $\pipeline$ algorithm uses the cycle property of the MST~\cite{tarjan}
to eliminate those inter-fragment edges that cannot
belong to the MST en route of their journey to the root. While the $\pipeline$ procedure, due to the pipelining of the edges to the root, takes $O(\sqrt{n})$
time (since there are at most so many MST edges left to be discovered after the end of the first part), it is not message-optimal: it
exchanges $O(m+ n^{1.5})$ messages, since each node in the BFS tree can send up to $O(\sqrt{n})$ edges leading to $O(n^{1.5})$ messages overall (the BFS tree construction takes $O(m)$ messages).

Our algorithm uses a different strategy to achieve optimality in both time and messages. 
The main novelty of our algorithm (Algorithm~\ref{alg:main}) is how the (at most) $\sqrt{n}$ base fragments
which remain at the end of the $\controlledGHS$ procedure are merge into one resulting fragment (the MST). 
Unlike previous time-optimal algorithms~\cite{kutten-domset,Elkin06,garay-sublinear}, we do not use the
$\pipeline$ procedure of~\cite{peleg-time-optimal,garay-sublinear}, since it is not message-optimal.
Instead, we continue to merge fragments, a la Bor\r{u}vka-style. 
%
Our algorithm uses two main ideas to implement the Bor\r{u}vka-style merging efficiently. (Merging is achieved by renaming the IDs of the merged fragments
to a common ID, i.e., all nodes in the combined fragment will have this common ID.) The first idea is a procedure to efficiently merge when $D$ is small (i.e., $D = O(\sqrt{n})$) or when the number of fragments remaining is small (i.e., $O(n/D)$).  The second idea is to use {\em sparse neighborhood covers} and efficient communication between fragments to merge fragments when $D$ is large {\em and} the number of fragments is large.
Accordingly, the second part of our algorithm can be divided into three phases, which are described next. %

\subsubsection{Phase 1: When $D$ is $O(\sqrt{n})$}
Phase 1 can be treated as a special case of Phase 3 (as in Algorithm~\ref{alg:main}).
However,  we describe Phase~1 separately as it helps in the understanding of the other phases as well.

We construct a BFS tree on the entire network, and perform the merging process as follows.
Each base fragment finds its LOE by convergecasting {\em within} each of its fragments. This takes $O(\sqrt{n})$ time and $O(\sqrt{n})$ messages per base fragment, leading to $O(n)$ messages overall. The $O(\sqrt{n})$ LOE edges  are sent 
by the leaders of the respective base fragments to the root by {\em upcasting} (see, e.g., \cite{peleg-locality}).
This takes $O(D+\sqrt{n})$ time and $O(D\sqrt{n})$ messages, as each of the $\sqrt{n}$ edges has to traverse up to $D$ edges
 on the way to the root.
 The root merges the fragments and sends the renamed fragment IDs to the respective leaders of the base fragments by {\em downcast} (which has the same time
 and message complexity as upcast~\cite{peleg-locality}). The leaders of the base fragments  broadcast the new ID to all other nodes in their respective fragments. This takes
 $O(\sqrt{n})$ messages per fragment and hence $O(n)$ messages overall.
  Thus one iteration of the merging can be done in $O(D+\sqrt{n})$ time and using $O(D\sqrt{n})$ messages.
 Since each iteration reduces the number of fragments by at least half, the number of iterations is $O(\log n)$. At the end of this iteration, several base fragments may  share the same label.  In subsequent iterations,  each base fragment  finds its LOE   (i.e., the LOE between itself and the other base fragments which do not have the same label) by convergecasting within its own fragment and (the leader of the base fragment) sends the LOE to the root; thus  $O(\sqrt{n})$ edges are sent to
 the root (one per base fragment), though there is a lesser number of combined fragments (with distinct labels).   The root  finds the overall LOE of the combined fragments and does the merging.
This is still fine, since the time and message complexity per merging iteration is $O(D+\sqrt{n})$  and $O(D\sqrt{n}) = O(n)$, respectively, as required.

\subsubsection{Phase 2: When $D$ and the Number of Fragments are Large}
When $D$ is large (say $n^{1/2+\epsilon}$, for some $0< \epsilon \leq 1/2$) and the number of fragments is large (say, $\Theta(\sqrt{n})$) the previous approach of merging via the root of the global BFS tree does not work directly, since the message complexity would be $O(D\sqrt{n})$.
The second idea  addresses  this issue:  we merge in a manner
that respects {\em locality}. That is, we merge fragments that are close by  using a {\em local} leader, such that the LOE edges do not have to travel too far.    
The high-level idea is to use a {\em hierarchy of  sparse neighborhood covers}  to accomplish the merging.\footnote{We use an efficient randomized cover construction algorithm due to Elkin~\cite{Elkin06}; this is the only randomization used  in our algorithm. Neighborhood covers were used by Elkin~\cite{Elkin06}  to 
improve the running time of the $\pipeline$ procedure of his distributed MST algorithm; on the other hand, here we use them to {\em replace} the
$\pipeline$ part entirely in order to achieve message optimality as well.}
A sparse neighborhood cover is a decomposition of a graph into a set of overlapping clusters
that satisfy suitable properties (see Definition~\ref{def:cover} in  Section \ref{sec:findpath}).  
The main intuitions behind using a cover are the following: (1) the clusters of the cover have relatively smaller  diameter (compared to the strong diameter of the fragment and is always bounded by $D$) and this allows efficient communication for fragments contained within a cluster (i.e., the weak diameter of the fragment is bounded by the cluster diameter); (2)
the clusters of a cover  overlap only a little, i.e., each vertex belongs only to a few clusters; this  allows essentially congestion-free (overhead is at most $\polylog(n)$ per vertex) communication and hence operations can be done efficiently in parallel across all the clusters  of a cover.
This phase continues till the number of fragments reduces to $O(n/D)$, when we switch to Phase 3. We next give more details on the merging process in Phase 2.
\medskip

\noindent {\bf Communication-Efficient Paths}.
An important technical aspect in the merging process is constructing
efficient communication paths between nearby fragments; the algorithm maintains and updates these efficient paths during the algorithm. Our algorithm requires fragments to be
``communication-efficient'', in the sense that there is an additional set of {\em short paths} between the
 fragment leader $f$ and fragment members. 
Such a path might use ``shortcuts'' through vertices in $V(G) \setminus V(F)$ to reduce the distance. 
The following definition formalizes this idea.

\begin{definition}[Communication-Efficient Fragment and Path]\label{def:efficientFrags}
Let $F$ be a fragment of $G$, and let $f \in F$ be a vertex designated as the \emph{fragment leader} of $F$.
We say that fragment $F$ is \emph{communication-efficient} if, for each vertex $v \in F$, there exists a path between $v$ and $f$ (possibly including vertices in $V(G) \setminus V(F)$) of length $O(\diam_G(F) +\sqrt{n})$, where $diam_G(F)$ is the  weak diameter of $F$.
Such a path is called \emph{communication-efficient path for $F$}.  
\end{definition}

Section~\ref{sec:routing} defines the routing data structures that are used to maintain communication-efficient paths.
Later, in \Cref{sec:findpath}, we describe the construction of  the paths (and routing data structures)  inductively.
We show that, in each iteration, all fragments find their respective LOEs in
time $\tilde O(\sqrt{n}+D)$ and using a total of $\tilde O(m)$ messages. While we cannot merge all fragments (along their LOEs),
as this will create long chains, we use a procedure called $\computeMM$ (Section \ref{sec:merge}) to merge fragments in a controlled manner. 
$\computeMM$ finds a maximal matching in the fragment graph $\cF_i$ induced by the LOE edges. The crucial part  is using communication-efficient paths to communicate efficiently (both
time and message-wise) between the fragment leader and the nodes in the fragment (while finding LOEs) as well as between fragment leaders
of adjacent fragments (while merging as well as implementing $\computeMM$). The procedure $\findLOE$ (see \Cref{sec:findLOE}) describes the LOE finding process assuming communication-efficient fragments. The maintenance   of such efficient fragments is shown recursively: the base fragments are efficient and  after merging the resulting fragments are  also efficient. 

We use a hierarchy of sparse neighborhood covers   to construct  communication-efficient  fragments (see \Cref{sec:findpath}).
Each cover in the hierarchy  consists of a collection of clusters of a certain radius: the lowest cover in the hierarchy has clusters of radius $O(\sqrt{n})$
(large enough to contain at least one base fragment, which has radius $O(\sqrt{n})$); subsequent covers in the hierarchy have clusters of geometrically increasing
radii, and the last cover in the hierarchy is simply the BFS tree of the entire graph. 
 Initially, it is easy to construct communication-efficient paths in base fragments, since they have strong diameter $O(\sqrt{n})$ (cf.\  Section \ref{sec:routing}, Lemma~\ref{lem:baseRouting}).  In subsequent iterations, when merging two adjacent fragments, the algorithm finds a cluster that is (just) large enough to contain both the fragments. 
\Cref{fig:fragments-clusters} gives an example of this process.
The neighborhood property of the cluster allows the algorithm to construct communication-efficient paths between merged fragments (that might take shortcuts outside the fragments, and hence have small {\em weak diameter}) assuming
that the fragments before merging are efficient.  
Note that it is important to make sure that the number of fragments in a cluster is not too large  in relation to the radius of the cluster---otherwise the message complexity would be high (as in the $\pipeline$ scenario).
Hence, a key invariant maintained through all the iterations is that the {\em cluster depth times the number of fragments that are contained in the cluster of such depth is always bounded by $\tilde{O}(n)$}, and this helps in keeping  the message complexity low. This invariant is maintained by making sure that the number of fragments per cluster {\em goes down} enough
to compensate for the increase in cluster radius (Lemma \ref{lem:invariant} in Section~\ref{sec:findpath}).
At the end of Phase~3, the invariant guarantees that when the cluster radius is $D$, the number of fragments is $O(n/D)$.

\subsubsection{Phase 3: When the Cluster Radius is $D$}

When the cluster radius becomes $D$ (i.e., the cover is just the BFS tree), we switch to Phase 3.
The number of remaining fragments will be $O(n/D)$ (which is guaranteed at the end of Phase 2). 
Phase 3 uses a merging procedure very similar to that of Phase 1. 
In Phase 1, in every merging iteration, each base fragment finds their respective LOEs (i.e., LOEs between itself and the rest of the fragments) by convergecasting to their respective leaders;  the   leaders  send at most $O(\sqrt{n})$ edges to the root by upcast. The root merges the fragments and sends out the merged information to the base fragment leaders by downcast.
In Phase 3, we treat the $O(n/D)$ remaining fragments as the ``base fragments''  and repeat the above process.
An important difference to Phase~1 is that the merging leaves the leaders of these base fragments intact: in the future iterations of Phase~3, each of these base fragments again tries to find an LOE using the procedure $\findLOE$, whereby only edges that have endpoints in fragments with distinct labels are considered as candidate for the LOE.

Note that the fragment leaders communicate with their respective nodes as well as the BFS root via the hierarchy of communication-efficient routing
paths  constructed  in Phase 2; these incur only a polylogarithmic overhead.
This takes $\tilde{O}(D+n/D)$ time (per merging iteration) since $O(n/D)$  LOE edges are sent to the root of the BFS tree via communication-efficient paths (in every merging iteration) and a message complexity of $\tilde{O}(D \cdot n/D) = \tilde O(n) $ (per merging iteration) since, in each iteration, each of the $O(n/D)$ edges takes $\tilde{O}(D)$ messages  to reach the root. Since there are $O(\log n)$ iterations overall, we obtain the desired bounds.

\section{Description and Analysis of the Algorithm}\label{sec:detailalgo}

\begin{algorithm*}[ht!]
\begin{algorithmic}[1]
  \footnotesize
\item[**] {\bf \large Part 1:} 
\STATE Run $\controlledGHS$ procedure (Algorithm \ref{alg:controlledGHS}).
\STATE Let $\cF_1$ be the base fragments obtained from $\controlledGHS$.
\item[**] {\bf \large Part 2:}
\item[*] \textbf{Start of Phase 1:}  
\FOR{every fragment $F \in \cF_1$}\label{line:phase1}
  \STATE Construct a BFS tree $T$ of $F$ rooted at the fragment leader.
  \STATE Every $u \in F$ sets $\up_u(F,1)$ to its BFS parent and $\down_u(F,1)$ to its BFS children.
\ENDFOR
\STATE Run the leader election algorithm of \cite{jacm15} to find a constant approximation of diameter $D$. %
\STATE \textbf{if} $D=O(\sqrt{n})$ \textbf{then} set $\cF' = \cF_1$ and skip to Phase~3 (Line~\ref{line:phase3}).
\item[*]\textbf{Start of Phase 2:} 
  \FOR[ All nodes start iteration $i$ at the same time]{$i=2,\dots,\lceil \log(D/\sqrt{n})\rceil$}\label{line:phase2}
  \STATE Construct cover $\cC_i = \cover(6 \cdot 2^{i+1} \cdot c_1 \sqrt{n})$, where $c_1$ is a suitably chosen constant. \label{line:computeCover}
  \STATE Every node locally remembers its incident edges of the directed trees in $\cC_i$. 
  \FOR{each fragment $F_1 \in V(\cF_i)$}
    \STATE Let $(u,v) = \findLOE(F_1)$ where $u \in F_1$ and $v \in F_2$. \COMMENT{$(u,v)$ is the LOE of $F_1$. See \Cref{sec:findLOE}.}
    \IF{$v \in F_2$ has an incoming lightest edge $e_1$ from $F_1$} \label{line:sym1}
      \STATE $v$ forwards $e_1$ to leader $f_2 \in F_2$ along its $((F_2,1),\dots,(F_2,i))$-upward-path. \label{line:sym2}
    \ENDIF
    \STATE $\findPath(F_1,F_2)$. \COMMENT{Find a communication-efficient path that connects leaders $f_1 \in F_1$ and $f_2 \in F_2$; this is needed for merging and also for iteration $i+1$. See \Cref{sec:findpath}.}
  \ENDFOR
  \item[] \COMMENT{{\bf Merging of fragments:}}
  \FOR{each fragment $F_1 \in V(\cF_i)$}
  \STATE \textbf{if} $F_1$ has a weak diameter of $\le 2^i c_1 \sqrt{n}$ \textbf{then} $F_1$ is marked active. \label{line:activef}
  \ENDFOR
  \STATE Let $\cM_i \subseteq \cF_i$ be the graph induced by the LOE edges whose vertices are the active fragments.
  \STATE Let $D$ be the edges output by running $\computeMM$ on $\cM_i$. \COMMENT{ We simulate inter-fragment communication using the communication-efficient paths.}
  \STATE \textbf{for} each edge $(F,F') \in D$: Mark fragment pair for merging. \label{line:mark1}
  \STATE \textbf{for} each active fragment $F$ not incident to an edge in $D$: Mark LOE of $F$ for merging.  \label{line:mark2}
  \STATE Orient all edges marked for merging from lower to higher fragment ID. A fragment leader whose fragment does not have an outgoing marked edge becomes \emph{dominator}.
  \STATE Every non-dominator fragment leader sends merge-request to its adjacent dominator.
  \FOR{each dominating leader $f$}
  \IF{leader $f$ received merge-requests from $F_1,\dots,F_\ell$} 
    \STATE Node $f$ is the leader of the merged fragment $F \cup F_1 \cup \dots \cup F_\ell$, where $F$ is $f$'s current fragment.
    \FOR{$j=1,\dots,\ell$}
      \STATE $f$ sends $\mu=\msg{\text{MergeWith},F}$ along its $(F_j,i)$-path to the leader $f_j$ of $F_j$.
      \STATE When $f_j$ receives $\mu$, it instructs all nodes $v \in F_j$ to update their fragment ID to $F$ and update all entries in $\up$ and $\down$ previously indexed with $F_j$, to be indexed with $F$.
    \ENDFOR
  \ENDIF
  \ENDFOR
  \STATE Let ${\cF}_{i+1}$ be the fragment graph consisting of the  merged fragments of $\cM_i$ and the inter-fragment edges.
\ENDFOR \label{line:mainEnd}
\item[] \textbf{end of iteration $i$.}
\STATE Let $\cF' = \cF_{\lceil\log(D/\sqrt{n})\rceil+1}$.
\item[*] \textbf{Start of Phase 3:} \COMMENT{Compute final MST given a fragment graph $\cF'$.} 
\FOR{$\Theta(\log n)$ iterations} \label{line:phase3}
\STATE Invoke $\findLOE(F')$ for each fragment $F' \in \cF'$ in parallel and then upcast the resulting LOE in a BFS tree of $G$ to a root $u$.  
\STATE Node $u$ receives the LOEs from all fragments in $\cF'$ and computes the merging locally. It then sends the merged labels to all the  fragment leaders
by downcast via the BFS tree.
\STATE Each fragment leader relays the new  label (if it was changed)  to all nodes in its own fragment via broadcast along the communication-efficient paths. 
\STATE At the end of this iteration, several fragments in $\cF'$ may share the same label. At the start of the next iteration, each fragment in $\cF'$  individually invokes $\findLOE$, whereby only edges that have endpoints in fragments with distinct labels are considered as candidates for the LOE.
\ENDFOR
\end{algorithmic}
\caption{\small{A Time- and Message-Optimal Distributed MST Algorithm.}}
\label{alg:main}
\end{algorithm*}\onlyLong{\par}

The algorithm operates on the \emph{MST forest}, which is a partition of the vertices of a graph into a collection of trees $\{T_1,\dots,T_\ell\}$
where every tree is a subgraph of the (final) MST. A \emph{fragment} $F_i$ is the subgraph induced by
$V(T_i)$ in $G$. We say that an MST forest is an \emph{$(\alpha,\beta)$-MST forest} if it contains at most
$\alpha$ fragments, each with a strong diameter\footnote{\label{ft:wd} Recall that the \emph{strong
diameter $\diam_F(F)$ of fragment $F$} refers to the longest shortest path (ignoring weights) between any two vertices in $F$
that only passes through vertices in $V(F)$, whereas the \emph{weak diameter} $\diam_G(F)$ allows the use
of vertices that are in $V(G) \setminus V(F)$.} of at most $\beta$.
Similarly, an MST forest is a \emph{weak $(\alpha,\beta)$-MST forest} if it contains at most $\alpha$ fragments each of (weak) diameter at most $\beta$.

We define the \emph{fragment graph}, a structure that is used throughout the algorithm.
The fragment graph $\cF_i$ consists of vertices $\{F_1,\dots,F_k\}$, where each $F_j$ ($1 \leq j \leq k$) is a fragment 
at the start of  iteration $i \geq 1$  of the algorithm.
The edges of $\cF_i$ are obtained by contracting the vertices of each $F_j \in V(\cF)$ to a single vertex in $G$ and removing all resulting self-loops of $G$.
We sometimes call the remaining edges \emph{inter-fragment} edges.
As our algorithm proceeds by finding lightest outgoing edges (LOEs) from each fragment, we operate partly on the \emph{LOE graph $\cM_i$ of iteration $i$}, which shares the same vertex set as $\cF_i$, i.e., $\cM_i \subseteq \cF_i$, but where we remove all inter-fragment edges except for one (unique) LOE per fragment.

\subsection{The $\controlledGHS$ Procedure}\label{sec:cghs}
Our algorithm starts out by making an invocation to the $\controlledGHS$ procedure introduced in~\cite{garay-sublinear}
and subsequently refined in~\cite{kutten-domset} and in~\cite{lenzenMSTlectureNotes}.

\begin{algorithm}[h]
\begin{algorithmic}[1]
\footnotesize
\STATE \textbf{procedure} $\controlledGHS$:
\STATE $\cF =  V(G)$ \COMMENT initial MST forest 
\FOR{$i=0,\dots,\lceil \log \sqrt{n} \rceil$}
\STATE $\cC =$ set of connectivity components of $\cF$ (i.e., maximal trees).
\STATE Each $C \in \cC$ of diameter at most $2^i$ determines the LOE of $C$ and add it to a candidate set $S$.
\STATE Add a maximal matching $S_M \subseteq S$ in the graph $(\cC,S)$ to $\cF$.
\STATE If $C \in \cC$ of diameter at most $2^i$ has no incident edge in $S_M$, it adds the edge it selected into $S$ to $\cF$.
\ENDFOR
\end{algorithmic}
\caption{Procedure~$\controlledGHS$: builds a $(\sqrt{n},O(\sqrt{n}))$-MST forest in the network.}\label{alg:controlledGHS}
\end{algorithm}

$\controlledGHS$ (Algorithm~\ref{alg:controlledGHS}) is a modified variant of the original GHS algorithm, whose purpose is to produce a balanced outcome
in terms of number and diameter of the resulting fragments (whereas the original GHS algorithm allows an
uncontrolled growth of fragments). This is achieved by computing, in each phase, a maximal matching
on the fragment forest, and merging fragments accordingly. Here we shall resort to the newest variant presented
in~\cite{lenzenMSTlectureNotes}, since it incurs a lower message complexity than the two preceding versions.
Each phase essentially reduces the number of fragments
by a factor of two, while not increasing the  diameter of any fragment by more than a factor of two. Since the number of phases
of $\controlledGHS$ is capped at $\lceil \log \sqrt{n} \rceil$,\footnote{Throughout, $\log$ denotes logarithm to the base 2.} it produces a $(\sqrt{n},O(\sqrt{n}))$-MST forest.
The fragments returned by the $\controlledGHS$ procedure are called \emph{base fragments}, and
we denote their set by $\cF_1$.

The following result about the $\controlledGHS$ procedure follows from~\cite{lenzenMSTlectureNotes}.

\begin{lemma}\label{lem:GHS}
Algorithm~\ref{alg:controlledGHS} outputs a $(\sqrt{n},O(\sqrt{n}))$-MST forest in $O(\sqrt{n} \log^* n)$
rounds and sends $O(m \log n + n \log^2 n)$ messages.
\end{lemma}
\onlyLong{
\begin{proof}
The correctness of the algorithm is established by Lemma~6.15 and Lemma~6.17 of~\cite{lenzenMSTlectureNotes}.
By Corollary~6.16 of~\cite{lenzenMSTlectureNotes}, the $i$-th iteration of the algorithm can be implemented in
time $O(2^i \log^* n)$. Hence the time complexity of $\controlledGHS$
is
\[
O\mleft(\sum_{i=0}^{\lceil \log \sqrt{n} \rceil} 2^i \log^* n \mright) = O\mleft(\sqrt{n} \log^* n \mright)
\]
rounds.

We now analyze the message complexity of the algorithm.
Consider any of the $\lceil \log \sqrt{n} \rceil$ iterations of the algorithm.
The message complexity for finding the lightest outgoing edge for each fragment (Line~5) is $O(m)$.
Then (Line~6) a maximal matching is built using the Cole-Vishkin symmetry-breaking algorithm.
As argued in the proof of~Corollary~6.16 of~\cite{lenzenMSTlectureNotes}, in every iteration
of this algorithm, only one message per fragment needs to be exchanged. Since the Cole-Vishkin algorithm
terminates in $O(\log^* n)$ iterations, the message complexity for building the maximal matching
is $O(n \log^* n)$. Afterwards, adding selected edges into $S$ to $\cF$ (Line~7) can be done
with an additional $O(n \log n)$ message complexity. The message complexity of algorithm
$\controlledGHS$ is therefore $O(m \log n + n \log^2 n)$.
\end{proof}
}

\subsection{Routing Data Structures for Communication-Efficient Paths}\label{sec:routing}
For achieving our complexity bounds, our algorithm maintains efficient fragments in each iteration.
To this end, nodes locally maintain routing tables.
In more detail, every node $u \in G$ has $2$ two-dimensional arrays $\up_u$ and $\down_u$ (called {\em routing} arrays), which are indexed by a (fragment ID,$\level$)-pair, where $\level$ stands for the iteration number, i.e., the for loop variable $i$ in Algorithm \ref{alg:main}.
Array $\up_u$ maps to one of the port numbers in $\{1,\dots,d_u\}$, where $d_u$ is the degree of $u$.
In contrast, array $\down_u$ maps to a {set} of port numbers. 
Intuitively speaking, $\up_u(F,i)$ refers to $u$'s parent on a path $p$ towards the leader of $F$ where  $i$ refers to the iteration in which this path was constructed.
Similarly, we can think of $\down_u(F,i)$ as the set of $u$'s children in all communication efficient paths originating at the leader of $F$ and going through $u$ and
we use $\down_u$ to disseminate information from the leader to the fragment members.
Oversimplifying, we can envision $\up_u$ and $\down_u$ as a way to keep track of the parent-child relations in a tree that is rooted at the fragment leader.  
(Note that $\level$ is an integer in the range $[1,\lceil \log(D/\sqrt{n})\rceil]$ that corresponds to the iteration number of the main loop in which this entry was added; see Lines~\ref{line:phase2}-\ref{line:mainEnd} of \Cref{alg:main}.)
For a fixed fragment $F$ and some value $\level=i$, we will show that the $\up$ and $\down$ arrays induce directed chains of incident edges.

Depending on whether we use array $\up$ or array $\down$ to route along a chain of edges, we call the chain an \emph{$(F,i)$-upward-path} or an \emph{$(F,i)$-downward-path}.
When we just want to emphasize the existence of a path between a node $v$ and a fragment leader $f$, we simply say that there is a  \emph{communication-efficient $(F,i)$-path between $v$ and $f$} and we omit ``$(F,i)$'' when it is not relevant.
We define the nodes specified by $\down_u(F,i)$ to be the \emph{$(F,i)$-children of $u$} and the node connected to port $\up_u(F,i)$ to be the \emph{$(F,i)$-parent of $u$}.
So far, we have only presented the definitions of our routing structures. We will explain their construction in more detail in \Cref{sec:findpath}.

We now describe the routing of messages in more detail:
Suppose that $u \in F$ generates a message $\mu$ that it wants to send to the leader of $F$. 
Then, $u$ encapsulates $\mu$ together with $F$'s ID, the value $\level=1$, and an indicator ``up'' in a message and sends it to its neighbor on port $\up_u(F,1)$; for simplicity, we use $F$ to denote both, the fragment and its ID.
When node $v$ receives $\mu$ with values $F$ and $\level=1$, it looks up $\up_v(F,1)$ and,
if $\up_v(F,1)=a$ for some integer $a$, then $v$ forwards the (encapsulated) message along the specified port.\footnote{Node $v$ is free to perform additional computations on the received messages as described by our algorithms, e.g., $v$ might aggregate simultaneously received messages in some form. Here we only focus on the forwarding mechanism.}
This means that $\mu$ is relayed to the root $w$ of the $(F,1)$-upward-path.
For node $w$, the value of $\up_w(F,1)$ is undefined and so $w$ attempts to lookup $\up_w(F,2)$ and then forwards $\mu$ along the $(F,2)$-upward-path and so forth. 
In a similar manner, $\mu$ is forwarded along the path segments $p_1 \dots p_i$ ($1\le j \le i$), where $p_j$ is the $(F,j)\text{-upward-path}$ in the $i$-th iteration of the algorithm's main-loop.
We will show that the root of the $(F,i)$-upward-path coincides with the fragment leader at the start of the $i$-th iteration.

On the other hand, when the iteration leader $u$ in the $i$-th iteration wants to disseminate a message $\mu$ to the fragment members, it sends $\mu$ to every port in the set $\down_u(F,i)$.
Similarly to above, this message is relayed to each leaf $v$ of each $(F,i)$-downward-path, for which the entry $\down_v(F,i)$ is undefined.
When $i>1$, node $v$ then forwards $\mu$ to the ports in $\down_v(F,j)$, for each $j<i$ for which $v$ is a root of the respective $(F,j)\text{-upward-path}$, and $\mu$ traverses the path segments $q_i \dots q_1$ where $q_\ell$ ($1 \le \ell \le i$) is the $(F,\ell)$-downward-path.
For convenience we call the concatenation of $q_i \dots q_1$ a \emph{$((F,i),\dots,(F,1))$-downward path}
(or simply \emph{$((F,i),\dots,(F,1))$-path}), and define a \emph{$((F,1),\dots,(F,i))$-upward path} similarly.

We are now ready to describe the individual components of our algorithm in more detail.
To simplify the presentation, we will discuss the details of \Cref{alg:main} inductively. 
We assume that every node $u \in F \in \cF_1$ knows its parent and children in a BFS tree rooted at the fragment leader $f \in F$.
(BFS trees for spanning each respective fragment can easily be constructed in $O(\sqrt{n})$ time and using a total of $O(m)$ messages---this is because the fragments in $\cF_1$ are disjoint and have strong diameter $O(\sqrt{n})$.)
Thus, node $u$ initializes its routing arrays by pointing $\up_u(F,1)$ to its BFS parent and by setting $\down_u(F,1)$ to the port values connecting its BFS children.

\begin{lemma}\label{lem:baseRouting}
At the start of the first iteration, for any fragment $F$ and every $u\in F$, there is an $(F,1)$-path between $F$'s fragment leader and $u$ with a path length of $O(\sqrt{n})$.
\end{lemma}
\onlyLong{
\begin{proof}
From the initialization of the routing tables $\up$ and $\down$ it is immediate that we reach the leader when starting at a node $u \in F$ and moving along the \emph{$(F,1)$-upward-path}.
Similarly, starting at the leader and moving along the \emph{$(F,1)$-downward-path}, allows us to reach any fragment member.
The bound on the path length follows from the strong diameter bound  of the base fragments, i.e., $O(\sqrt{n})$ (see Lemma~\ref{lem:GHS}).
\end{proof}
}

\subsection{Finding the Lightest Outgoing Edges (LOEs): Procedure~$\findLOE$}\label{sec:findLOE}
We now describe Procedure~$\findLOE(F)$, which enables  the fragment leader $f$ to obtain the lightest outgoing edge, i.e., the lightest edge that has exactly one endpoint in $F$.
Consider iteration $i\ge 1$.
As a first step, $\findLOE(F)$ requires all fragment nodes to exchange their fragment IDs with their neighbors to ensure that every node $v$ knows its set of incident outgoing edges $E_v$.
If a node $v$ is a leaf in the BFS trees of its base fragment, i.e., it does not have any $(F,1)$-children, it starts by sending the lightest edge in $E_v$ along the $((F,1),\dots,(F,i))$-upward-path.
In general, a node $u$ on an $(F,j)$-upward-path ($j\ge 1$) waits to receive the lightest-edge messages from all its $(F,j)$-children (or its $(F,j-1)$-children if any), and then forwards the lightest outgoing edge that it has seen to its parent in the $((F,j),\dots,(F,i))$-upward-path.

The following lemma proves some useful properties of $\findLOE$.
Note that we do not yet claim any bound on the message complexity at this point, as this requires us to inductively argue on the structure of the fragments, which relies on properties that we introduce in the subsequent sections.
Hence we postpone the message complexity analysis to Lemma~\ref{lem:findLOEMsgComplexity}.

\begin{lemma}[Efficient LOE Computation] \label{lem:findLOE}
Suppose that every fragment in $F \in \cF_i$ is communication-efficient at the start of iteration $i+1\ge 2$.
Then, the fragment leader of $F$ obtains the lightest outgoing edge by executing Procedure~$\findLOE(F)$ in $O(\sqrt{n} + \diam_G(F))$ rounds.
\end{lemma}
\onlyLong{
\begin{proof}
To accurately bound the congestion, we must consider the simultaneous invocations of $\findLOE$ for each fragment in $\cF_i$. 
Since, by assumption, every fragment is communication-efficient, every fragment node $u$ can relay its lightest outgoing edge information to the fragment leader along a path $p$ of length $O(\diam_G(F) + \sqrt{n})$.
Note that $p$ is precisely the $((F,1),\dots,(F,i))$-upward path to the leader starting at $u$.
To bound the congestion, we observe that the $(F,1)$-upward subpath of $p$ is confined to nodes in $F_u$ where $F_u$ is the base fragment that $u$ was part of after executing $\controlledGHS$.
As all base fragments are disjoint and lightest edge messages are aggregated within the same base fragment, the base fragment leader (who might \emph{not} be the leader of the current fragment $F$) accumulates this information from nodes in $F_u$ within $O(\sqrt{n})$ rounds (cf.\ Lemma~\ref{lem:baseRouting}).
After having traversed the $(F,1)$-upward path (i.e., the first segment of $p$) of each base fragment, the number of distinct messages carrying lightest edge information is reduced to $O(\sqrt{n})$ in total.
Hence, when forwarding any such message along a subsequent segment of $p$, i.e., an $(F_j)$-upward path for $j>1$, the maximum congestion at any node can be $O(\sqrt{n})$.
Using a standard upcast (see, e.g., \cite{peleg-locality})   and the fact that the length of path $p$ is $O( \diam_G(F) + \sqrt{n})$, it follows that the fragment leader receives all messages in $O(\diam_G(F) + \sqrt{n})$ rounds, as required.
\end{proof}
}

\subsection{Finding Communication-Efficient Paths: Procedure $\findPath$}\label{sec:findpath}

After executing $\findLOE(F_0)$, the leader $f_0$ of $F_0$ has obtained the identity of the lightest outgoing edge $e=(u,v)$ where $v$ is in some distinct fragment $F_1$.
Before invoking our next building block, Procedure~$\findPath(F_0,F_1)$, we need to ensure that both leaders are aware of $e$ and hence we instruct the node $v$ to forward $e$ along its $((F_1,1),\dots,(F_1,i))$-upward-path to its leader $f_1$ (see Lines~\ref{line:sym1}-\ref{line:sym2} of \Cref{alg:main}).

We now describe $\findPath(F_0,F_1)$ in detail. 
The goal is to compute a communication-efficient path between leaders $f_0$ and $f_1$ that can be used to route messages between nodes in this fragment. 
In \Cref{sec:merge}, we will see how to leverage these communication-efficient paths to efficiently merge fragments.

A crucial building block for finding an efficient path are the \emph{sparse neighborhood covers} that we precompute at the start of each iteration
(see Line~\ref{line:computeCover} of \Cref{alg:main}), and the properties of which we recall here. (Note that the cover definition  assumes
the underlying unweighted graph, i.e., all distances are just the hop distances.)

\begin{definition}\label{def:cover}
A \emph{sparse $(\kappa,W)$-neighborhood cover} of a graph is a collection $\cC$ of trees,
each called a {\em cluster}, with the following properties.
\begin{compactenum}
\item \emph{(Depth property)} For each tree $\tau \in \cC$, depth$(\tau) = O(W \cdot \kappa)$.
\item \emph{(Sparsity property)} Each vertex $v$ of the graph appears in $\tilde O(\kappa \cdot n^{1/\kappa})$ different trees $\tau \in \cC$.
\item \emph{(Neighborhood property)} For each vertex $v$  of the graph there exists a tree $\tau \in \cC$ that contains the entire $W$-neighborhood of vertex $v$.
\end{compactenum}
\end{definition}
Sparse neighborhood covers were introduced in~\cite{AP90}, and were found useful in several applications. 
We will use an efficient distributed (randomized) cover construction due to Elkin~\cite{Elkin06}\onlyLong{, which we recall here.\footnote{Although the algorithm as described in~\cite{Elkin06} is Monte Carlo, it can be easily converted to  Las Vegas.}}\onlyShort{, which we state in the full version as Theorem~1. See also Definition~1 in the full version.}

\begin{theorem}[{\cite[Theorem A.8]{Elkin06}}]\label{thm:cover}
There exists a distributed randomized Las Vegas algorithm, which here we call \cover, that constructs a
$(\kappa,W)$-neighborhood cover in time $O(\kappa^2\cdot n^{1/\kappa} \cdot\log n \cdot W)$
and using $O(m \cdot \kappa \cdot n^{1/\kappa} \cdot \log n)$ messages (both bounds hold with high probability) in the CONGEST model.
\end{theorem}

In our MST algorithm, we shall invoke Elkin's \cover procedure with $\kappa = \log n$, and write
\textsf{ComputeCover}($W$), where $W$ is the neighborhood parameter. 

\begin{figure}[t!]
    \begin{center}
       \includegraphics[width=0.65\textwidth]{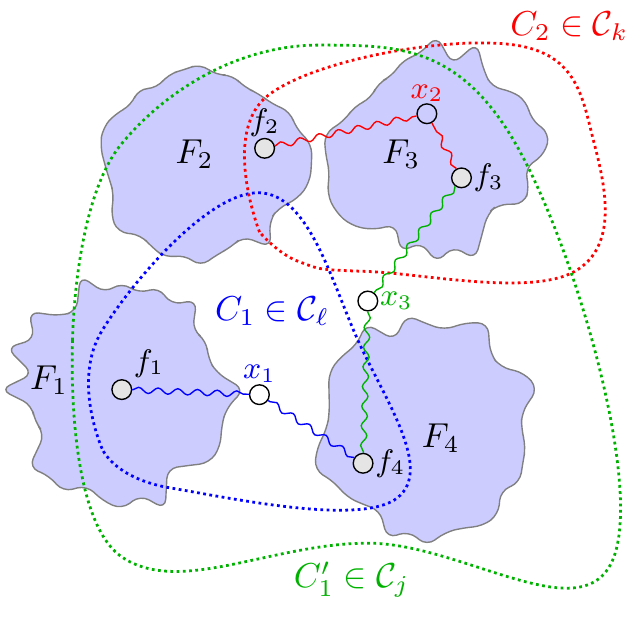}
       \caption{Fragments $F_1,\dots,F_4$. In the first iteration, $F_1,F_4$ and $F_2,F_3$ form adjacent fragment pairs that communicate along communication-efficient paths. $F_1$ and $F_4$ execute $\findPath$ and send probe messages along clusters of covers $\cC_1,\dots,\cC_\ell$ and finally succeed to find a communication-efficient path in a cluster $C_1 \in \cC_\ell$, which goes through the cluster leader $x_1 \in C_1$. Similarly $F_2$ and $F_3$ obtain a communication-efficient path in cluster $C_2 \in \cC_k$, after sending probe messages in clusters of covers $\cC_1,\dots,\cC_k$. In the next iteration, the merged  fragments $F_1 \cup F_4$ and $F_2 \cup F_3$ are (respectively) adjacent and proceed to construct a communication-efficient path in cluster $C_1' \in \cC_j$, after probing covers $\cC_1,\dots,\cC_j$.}
       \label{fig:fragments-clusters}
    \end{center}
\end{figure}

We are now ready to describe the communication-efficient paths construction.
As we want to keep the overall message complexity low, we start at the smallest cover construction $\cC_1$ and carefully probe for a cluster (tree) in $\cC_1$ that induces a communication-efficient path between $f_0$ and $f_1$.
Recall that every node locally keeps track of its incident cluster edges for each of the precomputed covers but we need to keep in mind that these structures are independent of the $\up$ and $\down$ arrays.
We instruct both leaders $f_0$ and $f_1$ to send a copy of their probe message to each of their $\cC_1$-parents. 
The parent nodes forward $u$'s probe message along their cluster tree to the root of their respective cluster tree.
Depending on whether a root receives the probe message in a timely fashion, we consider two cases:

\noindent\textbf{Case 1:} If there exists $C_w \in \cC_1$ such that $f_0, f_1 \in C_w$, then the probe message of both leaders reaches, through some path $p_0$ and $p_1$, the root $w \in C_w$ within $6 \cdot 2^2 c_1 \sqrt{n} \log n + O(\sqrt{n}\log^2 n)$ rounds, where the first term is $\radius(\cC_1)$ and the second term is to account for the congestion caused by simultaneous probe messages from the other fragment leaders (cf.\ Lemma~\ref{lem:findPath}). Then, $w$ replies by sending a ``success'' message back to $f_0$ and $f_1$ by reversing paths $p_0$ and $p_1$ to inform the leaders that they have found a communication-efficient path.

Note that it is possible for $f_0$ to receive multiple ``success'' reply messages.
However, since a cluster root only sends a success message if it receives probe messages from both leaders, $f_0$ and $f_1$ receive exactly the same set $M$ of success messages.
Thus they both pick the same success message sent by the cluster root node with the largest ID in $M$ (without loss of generality, assume that it is $w$) to identify the communication-efficient path and discard the other messages in $M$.

Suppose that $f_0$ received the message from $w$ along a path $p_0$ in cluster tree $C_w$.
Then, $f_0$ sends a message along $p_0$ and instructs every node $v$ in $p_0$ to set $\up_{v}(F_1,i+1)$ to the port of its successor (towards the root $w$) in $p_0$ and points $\up_{v}(F_0,i+1)$ to its predecessor in $p_0$.
When a node $v$ updates its $\up_v(F_1,i+1)$ array to some port $a$, it contacts the adjacent node $v'$ connected at this port who in turn updates $\down_{v'}(F_1,i+1)$ to point to $v$.
Similarly, leader $f_1$  and all nodes on the path $p_1$ proceeds updating their respective  $\up$ and $\down$ entries with the information provided by $p_1$ towards $w$.
Then, $f_0$ contacts its successor in $p_0$ to update its routing information whereas $f_1$ sends a similar request to its successor in $p_1$.
After these requests reach the cluster root $w$, the concatenated path $p_0\ p_1$ is a communication-efficient path between leaders $f_0$ and $f_1$.

\noindent\textbf{Case 2:}
On the other hand, if there is no appropriate cluster in $\cC_1$ that covers both leader nodes, then at least one of the two probe messages will arrive untimely at every cluster root and the leaders do not receive any success messages. 
Then, $f_0$ and $f_1$ rerun the probing process  by sending a probe message along their incident $\cC_2$ cluster edges and so forth. 
Note that all fragment leaders synchronize before executing the probing process.
We show in \Cref{lem:invariant} that all fragments have weak diameter at most $6 \cdot 2^{i} c_1 \sqrt{n}$ in iteration $i$.
Notice the radius of $\cC_i$ (see Line~\ref{line:computeCover}) ensures that $f_0$ and $f_1$ will arrive at a value $k \le i$, where $\cC_k$ is the cover having the smallest depth such that $f_0$ and $f_1$ are covered by some cluster in $\cC_k$ (but not by any cluster in $\cC_{k-1}$). Thus we can apply Case 1 for $\cC_k$.

Figure~\ref{fig:fragments-clusters} gives an example for the construction of communication-efficient paths. 

\begin{lemma}\label{obs:probeMsgs}
The number of probe messages that are generated by distinct fragment leaders and that are in transit simultaneously during an iteration of $\findPath$
is $O(\sqrt{n}\log^2 n)$ w.h.p.
\end{lemma}
\onlyLong{
\begin{proof}
Since, by Lemma~\ref{lem:GHS}, there are $O(\sqrt{n})$ base fragments, the total number of leaders at any point that are sending probe messages simultaneously is $O(\sqrt{n})$.
Note that, when exploring the communication efficient paths of a cover $\cC_j$, a leader needs to send a copy of its probe message to its parent in each of its $O(\log^2 n)$ clusters of $\cC_j$ that it is contained in. 
\end{proof}
}

\begin{lemma}\label{lem:invariant}
At the start of each iteration $i+1$, the fragment graph $\cF_i$ induces a weak $(\sqrt{n}/2^i, 6\cdot 2^i c_1 \sqrt{n}))$-MST forest in $G$.
\end{lemma}
\onlyLong{
\begin{proof}
We adapt the proof of Lemmas~6.15 and 6.17 of \cite{lenzenMSTlectureNotes} to show that
the fragment graph is a weak $(\sqrt{n}/2^i, 6\cdot 2^i c_1 \sqrt{n})$-MST forest.
For the case $i=1$, the claim follows directly from Lemma~\ref{lem:GHS}. We now focus on the inductive step $i>1$.

Suppose that $\cF_i$ is a weak $(\sqrt{n}/2^i, 6\cdot 2^i c_1\sqrt{n})$-MST forest.
We first argue that every new fragment in $\cF_{i+1}$ must have a weak diameter of at most $6 \cdot 2^{i+1} c_1 \sqrt{n}$. 

Consider the subgraph $M$ of $\cF_i$ induced by the edges marked for merging.
By Lines~\ref{line:mark1}-\ref{line:mark2} of Algorithm~\ref{alg:main}, each component of $M$ can contain at most one marked edge that was in the output of $\computeMM$. 
Thus, analogously to Lemma~6.15 in \cite{lenzenMSTlectureNotes}, it follows that
each component in $M$ contains at most one fragment of weak diameter $> 2^i c_1 \sqrt{n}$, since only fragments of weak diameter at most $2^i c_1 \sqrt{n}$ become active and participate in the matching.
Note that the maximality of the matching implies that each component of $M$ has diameter (in the fragment subgraph $M$) at most $3$. 
Moreover, all except at most $1$ fragment of such a component must have a weak diameter of at most $2^i c_1 \sqrt{n}$ since a fragment of a larger weak diameter does not select any edges for merging in this iteration.
It follows by the inductive hypothesis that the merged component has a weak diameter of at most $6 \cdot 2^i c_1 \sqrt{n} + 3 \cdot 2^i c_1 \sqrt{n} \le 6 \cdot 2^{i+1} c_1 \sqrt{n}$.

We now argue that each fragment contains at least $2^i c_2 \sqrt{n}$ nodes at the start of iteration $i>1$, assuming that it is true for all $j=1,\dots,i-1$.
To this end, consider the merging of fragments in iteration $i-1$. 
If a fragment $F \in \cF_i$ contains less than $2^i c_2 \sqrt{n}$ nodes it must have a weak diameter of at most $2^i c_2 \sqrt{n}$ and hence marks itself as active in Line~\ref{line:activef}. 
By the description of the merging process, $F$ is guaranteed to merge with at least one other fragment $F'$. 
By the inductive hypothesis, both $F$ and $F'$ consist of at least $2^{i-1} c_2 \sqrt{n}$ nodes and hence the merged fragment must have at least $2^i c_2 \sqrt{n}$ nodes, as required. 
\end{proof}
}

\begin{lemma}\label{lem:findPath}
  Consider any iteration $i\ge 1$.
  After the execution of $\findPath(F_0,F_1)$, there exists a communication-efficient path between leader $f_0$ and leader $f_1$ of length at most
  $O(2^k \sqrt{n})$, where $k\le i$ is the smallest integer such that there exists a cluster tree $C \in \cC_k$ such that $f_0, f_1 \in C$. 
  $\findPath(F_0,F_1)$ requires $O(2^k \sqrt{n}\log^2 n)$ messages and terminates in \[O\left(\sqrt{n}\log^2n + \min\{2^k \sqrt{n},\diam(G)\}\right)\] rounds with high probability.
\end{lemma}

\onlyLong{
\begin{proof}
By description of $\findPath$, leaders $f_0$ and $f_1$ both start sending a probe message along their incident $\cC_j$-edges towards the respective cluster roots, for $j=1,\dots,\lceil\log\sqrt{n}\rceil$.
First, note that $f_0$ and $f_1$ will not establish an efficient communication path for a cluster $C'$ in some $\cC_j$ ($j<k$), since, by definition, $f_0$ and $f_1$ are not both in $C'$ and hence one of the probe messages will not reach the root of $C'$.
To see that $k\le i$, note that \Cref{lem:invariant} tells us that in iteration $i$ every fragment has weak diameter at most $O(2^i\sqrt{n})$, whereas $\cC_i$ has a cluster radius of $\Theta(2^{i+1}\sqrt{n}\log n)$.

We now argue the message complexity bound.
Apart from the probe messages sent to discover the communication-efficient path in a cluster of cover $\cC_k$, we also need to account for the probe messages sent along cluster edges of covers $\cC_1,\dots,\cC_{k-1}$, thus generating at most
\begin{align*}
  \sum_{j=1}^{k}O(\radius(\cC_j)\log^2 n) &= \sum_{j=1}^k O(2^j \sqrt{n}\log^2 n) \\
  &\le 2^{k+1}O( \sqrt{n}\log^2 n) \\
  &= O(\radius(\cC_k)\log^2 n)
\end{align*}
messages, as required. 

Since $f_0$ and $f_1$ can communicate efficiently via a path $p$ leading through a cluster of cover $\cC_k$, then the length of $p$ is at most $2\,\radius(\cC_k)$. 
Applying Lemma~\ref{obs:probeMsgs} to take into account the additional congestion caused by simultaneous probe messages, yields a time complexity of $O(\radius(\cC_k) + \sqrt{n}\log^2 n)$. 
\end{proof}
}

\begin{lemma}\label{lem:findPathALL}
Consider an iteration $i$ and suppose that $\findPath$ is invoked simultaneously for each lightest outgoing edge.
Then, the total message complexity of all invocations is $O(n \log^3 n)$ and the time complexity is $\tilde{O}(\diam(G) + \sqrt{n})$ with high probability.
\end{lemma}

\onlyLong{
\begin{proof}
From Lemma~\ref{lem:invariant}, we know that every fragment in $\cF_i$ has weak diameter of $O(2^i \sqrt{n})$. 
Thus, every pair of adjacent fragments $F_0,F_1 \in \cF_i$ is covered by some cluster in cover $\cC_{i+1}$. 
In this case, Lemma~\ref{lem:findPath} tells us that a single invocation of $\findPath$ requires $O(2^{i+1}\sqrt{n} \log^2 n)$ messages.
Lemma~\ref{lem:invariant} tells us that there are $O(\sqrt{n}/2^i)$ fragments in $\cF_i$ (and thus also $O(\sqrt{n}/2^i)$ LOEs). 
Hence the total number of messages incurred by all pairs of fragments connected by an LOE is
\[
  O(2^{i+1}\sqrt{n} \log^2 n) \cdot O(\sqrt{n} / 2^i)  =  O(n \log^2 n).
\]
Summing up over all $i$, we obtain the claimed bound on the message complexity.

Finally we observe that Lemma~\ref{lem:findPath} already takes into account the congestion caused by simultaneous invocations, which yields the bound on the time complexity.  
\end{proof}
}

To summarize, Procedure $\findPath$ enables leaders of adjacent fragments to communicate with each other by sending messages along the communication-efficient paths given by the routing tables $\up$ and $\down$.

\subsection{Merging Fragments}\label{sec:merge}
We will avoid long chains of merged fragments by using procedure $\computeMM$~\cite{lenzenMSTlectureNotes}.
$\computeMM$ outputs a maximal matching on a fragment forest, where fragments in $\cF_i$ are treated as
super-vertices of a graph connected by inter-fragment edges.
Procedure $\computeMM$ simulates  the Cole-Vishkin symmetry-breaking distributed algorithm, which terminates in
$O(\log^* n)$ iterations~\cite[Theorem 1.7]{lenzenMSTlectureNotes}. We next show how to do the simulation efficiently in the fragment graph.

Procedure $\findPath$ enables communication via communication-efficient paths between any two adjacent fragment leaders in $\cM_i$.
In turn, this enables the simulation of procedure $\computeMM$ on the network induced by $\cM_i$, where the leaders in $\cM_i$ perform the computation required by $\computeMM$.
The following lemma follows directly from Lemma~\ref{lem:findPathALL}.

\begin{lemma}\label{cor:simulation}
Suppose that every fragment in $\cF_i$ is efficient and let $\cM_i\subset \cF_i$ be the lightest outgoing edge graph obtained by running $\findPath$.
Then, $\computeMM$ can be simulated on the network defined by $\cM_i$, requiring $\tilde{O}(\diam(G) + \sqrt{n})$ rounds and $\tilde{O}(n)$ messages. 
\end{lemma}

Every non-dominator fragment $F_1'$  sends a $\msg{\text{MergeReq}}$ message to the leader $f_1'$ of an arbitrarily chosen adjacent dominator fragment $F$. 
The dominator fragment processes all merge-requests in parallel and replies by sending a $\msg{\text{MergeWith},F}$ message to the leader $f'$ of each fragment $F'$ from which it received $\msg{\text{MergeReq}}$; in turn, $f'$ forwards this request along the $((F',i),\dots,(F',1))$-downward path to every node in $F'$.
Upon receiving a $\msg{\text{MergeWith},F}$ message, node $u' \in F'$ updates its fragment ID to $F$, and also updates its routing table by setting $\up_{u'}(F,\ell) = \up_{u'}(F',\ell)$ and $\down_{u'}(F,\ell) = \down_{u'}(F',\ell)$, for every value of $\ell$.
Note that the leader of the dominator fragment becomes the new leader of the merged fragment.

\begin{lemma}\label{lem:merge}
Consider iteration $i$.
If, for each $j\le i$, every fragment in $\cF_j$ is communication-efficient, then the following hold.
\begin{compactenum}
\item With high probability, the message complexity for merging fragments in iteration $i$ is $\tilde O(m)$ and the process completes within $\tilde O(\diam(G) + \sqrt{n})$ rounds.
\item Every fragment in $\cF_{i+1}$ is communication-efficient.
\end{compactenum}
\end{lemma}

\onlyLong{
\begin{proof}
To show (1), we argue recursively starting at iteration $i$, as follows:
note that forwarding the $\msg{\text{MergeWith}}$ and $\msg{\text{MergeReq}}$ messages requires communicating between neighboring fragments and thus by Lemma~\ref{cor:simulation} we require $O(\diam(G) + \sqrt{n})$ rounds and $O(n\log^2 n)$ messages.
Consider an adjacent pair of fragments $F_0$ and $F_1$ and suppose that $F_0$ merges with the dominator fragment $F_1$.
Since we eventually need to broadcast the new fragment ID to every node $u \in F_0$ we need to ensure that the routing tables $\up_u(F_1,\cdot)$ and $\down_u(F_1,\cdot)$ are updated correctly to route messages towards the new leader $f_1 \in F_1$ (and vice versa from $f_1$ to all nodes in $F_1$), when we compute the lightest outgoing edge of the merged fragment $F_0 \cup F_1$ in subsequent iterations.
If $i>1$, then $F_0$ might be composed of merged fragments $F_0' \cup \dots \cup F_\ell'$ that merged  in previous iterations; without loss of generality, suppose that this iteration is $i-1$. 
By assumption, $\cF_{i-1}$ consisted of efficient fragments.
As nodes do not remove routing information from $\up$ and $\down$, the leader $f_0$ can use the communication-efficient paths obtained by invoking $\findPath$ in iteration $i-1$ to forward the new fragment ID to the leaders of the $F_0',\dots,F_\ell'$, which we call the \emph{$(i-1)$-iteration fragments}.
Applying Lemma~\ref{cor:simulation} to $\cM_{i-1}$ reveals that we can use the paths obtained by invoking $\findPath$ in iteration $i-1$ to relay the new fragment ID to $(i-1)$-iteration fragments while incurring only $O(\diam(G) + \sqrt{n})$ rounds and $O(n\log^2 n)$ messages in total.
Recursively applying this argument until iteration $1$, allows us to reason that 
$O((\diam(G) + \sqrt{n})\log n)$ rounds and $O(n\log^3 n)$ messages are sufficient to relay all new fragment IDs to the base fragment leaders.
At this point, every base fragment leader uses the BFS tree of the base fragments to broadcast this information to the base fragment nodes, requiring $O(\sqrt{n})$ rounds and $O(m)$ messages.

To show (2), we observe that $\cF_i$ consists of communication-efficient fragments, and hence every fragment node $u \in F_j$ of a newly merged fragment $F = F_1 \cup \cdots \cup F_\ell$ ($\ell \ge j$) can already communicate efficiently with the leader $f_j$ in its subfragment $F_j$, which has now become part of $F$.
Moreover, the paths obtained by $\findPath$ ensure that $f_j$ can communicate efficiently with leader $f \in F$ and hence it follows transitively that $u$ has a communication-efficient path to $f$, as required.
\end{proof}
}
The analysis of the message complexity of merging fragments allows us to obtain a bound on the number of messages required for computing a lightest outgoing edge in each fragment.

\begin{lemma}\label{lem:findLOEMsgComplexity}
The message complexity of all parallel invocations of $\findLOE$ is $\tilde O(m)$ in total w.h.p.
\end{lemma}

\onlyLong{
\begin{proof}
In the first step of $\findLOE$, each node exchanges messages with its neighbors requiring $\Theta(m)$ messages. 
Let $F = F_1\cup\dots\cup F_\ell$, where $F_1,\dots,F_\ell$ are base fragments, and consider some vertex $u \in F_1$. 
As previously argued, $u$ relays its LOE information along the $((F,1),\dots,(F,i))$-upward-path to the fragment leader and the segment formed by the $(F,1)$-upward path ends at the base fragment leader of $F_1$, which are exactly the BFS trees yielded by \controlledGHS.
A crucial observation is that $u$ only sends its LOE information to its parent in the path, \emph{after} receiving the LOE messages from all its children (see \Cref{sec:findLOE}).
This ensures that each node sends exactly one message and hence we obtain a bound of $\sum_{j=1}^\ell O(|V(F_j)|) = O(|V(F)|)$ on the number of messages sent in the $(F,1)$-upward-path of the nodes in $F$. 
This is subsumed in the message complexity of exchanging messages with neighbors in the first step, which is $O(m)$.

At this point, each base fragment leader $f_j$ of $F_j$ ($j=1,\dots,\ell$) holds exactly one (aggregated) lightest outgoing edge information message $\mu_j$, which needs to be relayed to the fragment leader $f$ of $F$ along the respective $((F,2),\dots,(F,i))$-upward-path of $ O(\diam_G(F))$ hops (see Definition~\ref{def:efficientFrags}).

By reversing the argument used for proving part~(2) of Lemma~\ref{lem:merge}, we can inductively apply Lemma~\ref{cor:simulation} to obtain a bound of  $O(n\log^3 n)$ messages per iteration, and thus the total message complexity is $O(m + n\log^3 n) = \tilde O(m)$.
\end{proof}
}

\begin{lemma} \label{lem:phase3}
Phase~3 of the algorithm requires $\tilde O(m)$ messages and $\tilde O(D + \sqrt{n})$ time and ensures that all fragments have the same label (i.e., are merged).
\end{lemma}
\begin{proof}
Note that our algorithm either executes Phase~3 directly after Phase~1 (thus skipping Phase~2) or after executing Phase~2. 
First we argue (for both cases) that all fragments have the same fragment ID after the $\Theta(\log n)$ iterations in Phase~3.
To see that the number of fragment labels is at least halved in each iteration, note that, when executing $\findLOE$, all nodes exchange their fragment IDs with their neighbors (requiring $O(m)$ messages) and then only choose candidate LOE edges that have their endpoint in fragments with distinct IDs.
This ensures that every fragment pairs up with another fragment and hence one of the two distinct IDs will be removed; note that long ``chains'' of fragments connected by LOE edges are possible and result in an even faster reduction of distinct labels---all fragments in the chain adapt the root fragment ID (cf.\ Phase~3 in the pseudo code). 
Thus, after the last iteration of Phase~3, all fragments carry the same fragment ID and no more LOE edges are required as all fragments are considered to be merged. 

Now we consider the message and time complexity of Phase~3.
According to Lemma~\ref{lem:findLOE}, the time complexity of finding the LOEs is $O(D+ \sqrt{n})$, and according to Lemma~\ref{lem:findLOEMsgComplexity} $\tilde O(m)$ messages are required to find the LOEs.
This is true independently of whether we called Phase~3 directly after Phase~1 or after Phase~2.

Now, consider the case where we execute Phase~3 directly after Phase~1 (thus skipping Phase~2), i.e., $D = O(\sqrt{n})$. 
Here, $\findLOE$ results in each node locally determining the incident LOE and then aggregating the LOE to the base fragment leader.
In addition to the base fragment BFS trees, we also construct a global BFS tree $T$, which, has $O(\sqrt{n})$ diameter by assumption.
The base fragment leaders then forward their respective LOE along towards the root $u$ of $T$. Since we have $O(\sqrt{n})$ distinct base fragments, there are at most $O(\sqrt{n})$ LOE edges sent upward in $T$, thus resulting in an additional message complexity of $O(D\sqrt{n}) = O(n)$. Taking into account that it takes $O(\sqrt{n})$ rounds for the base fragment leaders to determine the LOE of their fragment, the time complexity amounts to $O(D + \sqrt{n})$. 

We now argue the message and time complexity for the case where we execute Phase~3 after Phase~2. 
Here, we start out with $O(n/D)$ distinct fragments each having their own fragment ID and a global BFS tree $T$ of depth $O(D)$. 
Since each fragment finds $1$ LOE which is first aggregated at the fragment leader and then forwarded along $T$ to the global BFS root,
this requires $O(D \cdot n/D) = O(n)$ messages in total and $O(D + n/D) = O(D)$ rounds, since $D = \Omega(\sqrt{n})$ by assumption, completing the proof.
\end{proof}

Combining the complexity bounds from the previous lemmas we obtain the following theorem.

\begin{theorem}\label{thm:main}
Consider a synchronous network (in the $KT_0$ model) of $n$ nodes, $m$ edges, and diameter $D$,
and suppose that at most $O(\log n)$ bits can be transmitted over each link in every round.
Algorithm~\ref{alg:main} computes an MST and, with high probability, runs in $\tilde O(D + \sqrt{n})$ rounds and exchanges $\tilde O(m)$ messages.
\end{theorem}

\onlyLong{
\section{A Simultaneously Tight Lower Bound}\label{sec:lowerbound}
As mentioned in Section~\ref{sec:result}, the existing graph construction of \cite{elkin,stoc11} used to establish
the lower bound of $\tilde\Omega(D+\sqrt{n})$ rounds does not simultaneously yield the message lower bound of $\Omega(m)$;
similarly, the existing lower bound graph construction of \cite{jacm15} that shows the message lower bound
of $\Omega(m)$ does not simultaneously yield the time lower bound of $\tilde\Omega(D + \sqrt{n})$.
Previously, \cite{stoc11} presented a sparse graph of $O(n)$ edges to obtain the $\tilde\Omega(D + \sqrt{n})$ time bound for almost all choices of $D$, while \cite{jacm15} showed that $\Omega(m)$ messages are required to solve broadcast and hence also for constructing a (minimum) spanning tree.\footnote{Any algorithm that constructs an spanning tree using $O(f(n))$ messages can be used to elect a leader using $O(f(n) + n)$ messages in total, by first constructing a spanning tree and then executing any broadcast algorithm  restricting its communication to the $O(n)$ spanning tree edges.}

The following result presents a ``universal lower bound'' for MST in the sense that it shows that for essentially any $n$, $m$, and $D$, there exists a class of graphs of $n$ nodes, $m$ edges, and with diameter $D$, for which every randomized MST algorithm takes $\tilde\Omega(D + \sqrt{n})$ rounds {\em and} $\Omega(m)$ messages to succeed with constant probability. Our proof combines two lower bound techniques: hardness of distributed symmetry breaking, used to show the lower bound on message complexity \cite{jacm15}, and communication complexity, used to show the lower bound on time complexity~\cite{stoc11}.

\begin{figure}[!ht]
    \begin{center}
       \includegraphics[width=0.98\textwidth]{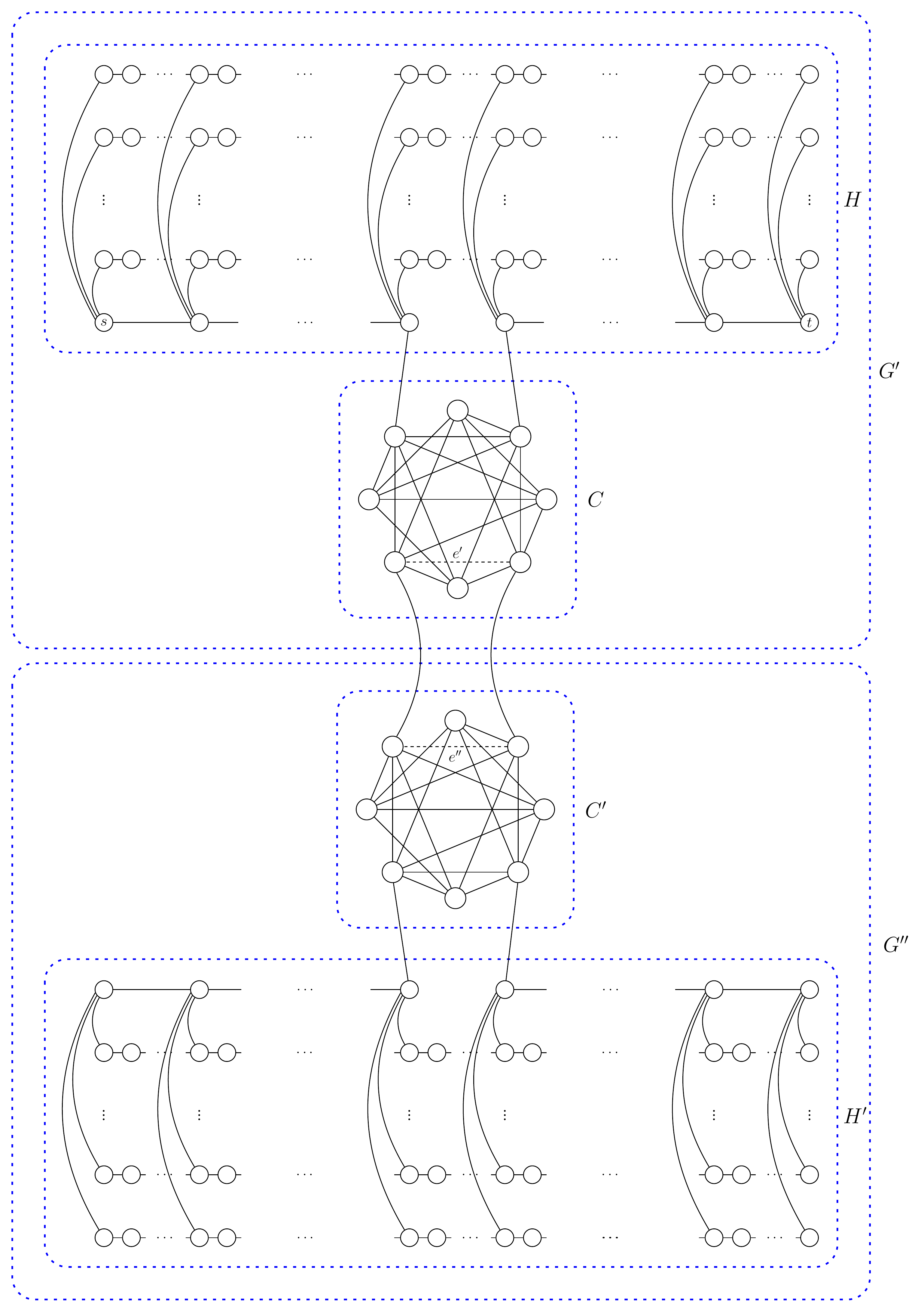}
       \caption{The graph $Dumbbell(G'[e'],G''[e''])$ for the proof of \Cref{thm:lb}.}
       \label{fig:lowerbound}
    \end{center}
\end{figure}

\begin{theorem}\label{thm:lb}
There is a class of graphs of $n$ nodes, $m$ edges (for $ n \leq m \leq {n \choose 2}$), and diameter $D = \Omega(\log n)$ for which every $\epsilon$-error distributed MST algorithm requires $\Omega(m)$ messages and $\tilde\Omega(D + \sqrt{n})$ time in expectation in the $KT_0$ model, for any sufficiently small constant $\epsilon>0$. 
This holds even if nodes have unique IDs and have knowledge of the network size $n$.
\end{theorem}
}

\subsection{Proof of \Cref{thm:lb}}

\paragraph{The Lower Bound Graph}
Our lower bound graph $G$ consists of the graph construction $H$ of \cite{peleg-bound} (and its subsequent refinement in \cite{stoc11}), combined with the \emph{dumbbell graph} construction of~\cite{jacm15}.
We first outline the main features of $H$, and refer the reader to \cite{stoc11} for the details. 
The graph $H$ consists of two designated nodes $s$ and $t$ that are connected by $\Theta(\sqrt{n})$ vertex-disjoint \emph{slow paths}, each having length $\Theta(\sqrt{n})$ and one \emph{highway path} of length $D$, which determines the diameter of $H$ by adding spokes (i.e., shortcuts) at appropriate points to the slow paths.

We adapt this graph by removing the edge between the two vertices $u_1$ and $u_2$ on the highway path at distance $\lfloor D/2 \rfloor$ and $\lceil D/2 \rceil + 1$ from $s$ and connecting them to one vertex each of a $\lceil c_1 m/n \rceil $-regular graph $C$ consisting of $c_2 n$ nodes, where $c_1$ and $c_2$ are two positive constants.
We assume that $C$ has a strong diameter of $O(\log n)$, where $m \ge c n$, for a sufficiently large positive constant $c$.\footnote{Such graphs exist since any random $d$-regular graph is known to be an expander (and hence its diameter is $O(\log n)$) with high probability when $d$ is sufficiently large (at least some constant).} 
We call the edges of $C$ \emph{switch edges}.
Note that the two vertices of $C$ that are connected to $u_1$ and $u_2$ have degree $\lceil c_1 m/n \rceil +1$.

To obtain a concrete graph from the lower bound construction, we assign unique IDs (chosen from a range of size $\poly(n)$), and specify a port mapping for each node $u$ that maps $[1,\text{deg}(u)]$ to one of $u$'s neighbors. We point out that this port mapping function is not known in advance to $u$.
For a concrete graph $G$, we define the \emph{open graph $G[e]$} as the graph where we have removed edge $e$, and we define $\G^{\text{open}}$ to be the set of open graphs obtained by all possible ways of removing any of the switch edges in $C$.
Note that this is different from the construction in \cite{jacm15}, where $\G^{\text{open}}$ consists of all open graphs considering \emph{all} possible edge removals. 
Let $G'[e'], G''[e''] \in \G^{\text{open}}$ be two open graphs with disjoint node IDs.
By connecting the two open ports (due to removing edge $e'$) of $G'$ to the two open ports in $G''$ we obtain the graph
$Dumbbell(G'[e'],G''[e''])$. These two new edges are called \emph{bridge edges}. See Figure~\ref{fig:lowerbound}.

\subsubsection{Part 1: Symmetry Breaking}

\paragraph{The Complexity of Bridge Crossing and Broadcast}
We define the \emph{input graph collection} $\I$ to be the set of all dumbbell graphs obtained by bridging ID-disjoint open graphs from $\G^{\text{open}}$, which contains all possible $1$-edge removals of all possible concrete graphs taking into account all possible port numberings and ID assignments.

To solve the \emph{bridge crossing problem} on a graph in $\I$ (we refer to~\cite{jacm15} for its definition), an algorithm is required
to send a message across one of the two bridge edges.
By~\cite[Lemma~3.11]{jacm15}, any deterministic algorithm that solves broadcast on a constant fraction of the inputs (assuming a uniform distribution) must also solve the bridge crossing problem with an expected message complexity of $\Omega(m)$, assuming that inputs are sampled uniformly from $\I$. The result is extended to randomized Monte Carlo algorithms with constant error probability by Yao's Minimax
Lemma~\cite{yaominimax}.
We cannot apply this result directly to our setting, as our set $\I$ is restricted to all possible dumbbell graph combinations for the switch edges in $C$ rather than considering all edges of the graph. 
Nevertheless, the number of switch edges in $C$ is $|E(C)| = \Theta(\tfrac{m}{2n}|C|) = \Theta(m) = \Theta(|E(G)|)$ and hence the counting argument of Lemmas~3.5 and 3.6 in \cite{jacm15} can be adapted to show an average message complexity of $\Omega(m)$ for solving bridge crossing with a deterministic algorithm when choosing input graphs uniformly from $\I$. We sketch the argument and refer the reader to \cite{jacm15} for the details: The main idea of the proof is to consider $Dumbbell(G'[e'],G''[e''])$, where each of $G'$ and $G''$ is a copy of $G$ with a concrete port numbering and ID assignment.
Let $\cP$ be a bridge crossing algorithm and consider the execution of $\cP$ on the (disconnected) graph consisting of $G'$ and $G''$ of $2n$ nodes.
Comparing this with the execution of $\cP$ on $Dumbbell(G'[e'],G''[e''])$, an easy indistinguishability argument shows that $\cP$ behaves exactly the same in both executions up until the point where bridge crossing happens. 
In the execution on the disconnected graph let $t(e)$ be the first time that $\cP$ sends a message across $e$, for any $e \in E(C)$, and let $L=(e_1,\dots,e_\ell)$ be a list containing the edges of $G'$ in increasing order of $t(e)$, breaking ties in a predetermined way.
It follows that, when $\cP$ sends the first message across $e_j$ on $Dumbbell(G'[e'],G''[e''])$, which occurs at the $j$-th position in $L$, it must have sent at least $j-1$ messages for $e_1,\dots,e_{j-1}$. We obtain the average message complexity for deterministic algorithms by counting the total number of messages in all graphs in $\I$ divided by the number of graphs in the input collection (see Lemma~3.5 in \cite{jacm15}). This extends to randomized Monte Carlo algorithms via Yao's Lemma \cite{yaominimax}; thus, we have the following result.

\begin{lemma}\label{lem:bridgecrossing}
Let $\cP$ be an $\epsilon$-error randomized broadcast algorithm.
Then, there is a graph $G \in \I$ such that the expected message complexity of $\cP$ on $G$ is $\Omega(m)$, where the expectation is taken over the random bits of $\cP$.
\end{lemma}

\subsubsection{Part 2: Communication Complexity}

\paragraph{Reduction from Set Disjointness} The lower bound for MST of \cite{stoc11} is shown by a reduction from the spanning connected subgraph problem, which itself is used in a reduction from the set disjointness problem in 2-party communication complexity~\cite{KushilevitzN97}. In the two party model,  Alice receives $X$ and Bob receives $Y$, for some $b$-bit vectors $X$ and $Y$, and the players communicate along a communication channel to decide if there is an index $i$ such that $X[i] = Y[i] = 1$.
Razborov~\cite{Razborov92} showed that any $\epsilon$-error randomized error communication protocol requires $\Omega(b)$ bits to solve set disjointness. \cite{stoc11} showed how Alice and Bob can jointly simulate the execution of a distributed MST algorithm $\cA$ in the graph $G$ with a weight assignment depending on the inputs $X$ and $Y$  to obtain a protocol for set disjointness as follows:
All slow path edges and all highway edges obtain weight $1$ in $G$, whereas the spoke edges that are not incident to $s$ or $t$ obtain weight $\infty$.
We assign weight $1$ to all edges in $E(C)$.
For every $i\in [1,b]$, the $i$-th spoke edge incident to $s$ is assigned weight $1$ if $X[i]=0$, and weight $n$ otherwise.  
Similarly, the $i$-th spoke edge incident to $t$ is assigned weight $1$ if $Y[i]=0$, and weight $n$ otherwise.  
Consider the MST $M$ of $G$ the $j$-th slow path $\rho_j$ connecting $s$ and $t$. 
A crucial property is that $\rho_j$ must contain exactly one spoke incident to either $s$ or $t$ as otherwise $\rho_j$ is either disconnected from the rest of the graph or, if both spokes are part of $M$, the highway path forms a cycle with $\rho_j$. 
If $X$ and $Y$ are disjoint, then either the $j$-th spoke incident to Alice has weight $1$ or the $j$-th spoke incident to Bob; in this case, the spoke that has weight $1$ is part of $M$. As a consequence, the MST contains one edge of weight $n$ if and only if $X$ and $Y$ are not disjoint.

\paragraph{Simulating the MST algorithm}
Alice and Bob create $G$, assign weights appropriately to the edges incident to $s$ and $t$, and then simulate the execution of $\cA$ on $G$; essentially, Alice simulates all nodes except $t$ and its neighbors and, similarly, Bob simulates all nodes except $s$ and its neighbors. To keep the simulation of $s$ and $t$ afloat, Alice and Bob exchange at most $2$ bits per simulated round; we refer the reader to \cite{stoc11} for the details.
Once $\cA$ terminates, Alice knows which edges incident to $s$ are in the MST and Bob knows the same about $t$.
Moreover, since the weight of the MST depends only on these incident edges, Alice can compute the total weight incident to $s$ and then send it to Bob, requiring $O(\log n)$ bits. From this, Bob can reconstruct the total weight of the MST (since all other edges have weight $1$).
If the MST does not contain any edge of weight $n$, then the total weight is $n-1$ and, by the above correspondence,
Bob can conclude that $X$ and $Y$ are disjoint.
On the other hand, if the MST does contain an edge of weight $n$ (which must be a spoke incident to either $s$ or $t$) then there is some index where $X$ and $Y$ intersect.
It follows that the solution for MST solves set disjointness and it follows that the simulation cannot terminate in $o(b)$ rounds as this will result in $o(b)$ bits being communicated between Alice and Bob, contradicting the $\Omega(b)$ lower bound for set disjointness~\cite{Razborov92}. 
Since this holds for a constant probability of error, an easy application of Markov's inequality shows that the expected time complexity must also be $\Omega(D + \sqrt{n})$:

\begin{lemma}\label{lem:cc}
There exists a weight function $w$ such that, for any graph $G \in \I$, executing algorithm $\cA$ on the weighted graph $G_w$, where every edge $e$ has weight $w(e)$, 
takes $\Omega(D + \sqrt{n})$ rounds in expectation.
\end{lemma}

\subsubsection{Putting Everything Together}

We are now ready to combine the results of \Cref{lem:bridgecrossing}, which we only argued for unweighted graphs so far, with \Cref{lem:cc}.
The next lemma directly implies \Cref{thm:lb}.

\begin{lemma} \label{lem:combining}
There exists a weighted graph $G$ such that any MST algorithm requires $\tilde \Omega(\sqrt{n} + D)$ rounds in expectation and has an expected message complexity of $\Omega(m)$.
\end{lemma}

\begin{proof}
Consider an MST algorithm $\cA$ and the worst case weight assignment $w$ provided by \Cref{lem:cc}.
Apply $w$ to every graph in the collection $\I$ yielding the collection of weighted graphs $\bar{\I}$. 
Note that, for any $G \in \I$, the edges of the corresponding weighted graph $\bar{G} \in \bar{\I}$ have weight $1$.
Thus we can apply the arguments preceding \Cref{lem:bridgecrossing} to the weighted graph collection $\bar{\I}$ to obtain the result.
\end{proof}

\section{Conclusions}
We have presented a new distributed algorithm for the fundamental minimum spanning tree problem which is
simultaneously time- and message-optimal (to within $\polylog(n)$ factors). 

An interesting open question is whether there exists a distributed MST algorithm with
near-optimal time and message complexities in the $KT_1$ variant of the model.

Currently, it is not known whether other important problems such as shortest paths, minimum cut,
and random walks, enjoy singular optimality. These problems admit distributed algorithms which are
(essentially) time-optimal but not message-optimal~\cite{danupon1,danupon2,jacm13,podc11}.
Further work is needed to address these questions.

\end{document}